\documentclass[conference]{IEEEtran}
\IEEEoverridecommandlockouts
\usepackage{cite}
\usepackage{amsmath,amssymb,amsfonts}
\usepackage{algorithmic}
\usepackage{graphicx}
\usepackage{textcomp}
\usepackage{xcolor}
\usepackage{verbatim}

\usepackage{algorithm}

\usepackage[english]{babel}
\usepackage[labelfont=bf]{caption}
\usepackage[pagewise]{lineno}
\usepackage{paralist}
\usepackage{enumitem}
\usepackage{varwidth}
\usepackage{caption,subcaption}

\usepackage{multirow}

\newtheorem{myDef}{Definition}
\newtheorem{property}{Property}

\def\BibTeX{{\rm B\kern-.05em{\sc i\kern-.025em b}\kern-.08em
    T\kern-.1667em\lower.7ex\hbox{E}\kern-.125emX}}
\begin{document}

\title{\texttt{Tao}: A Learning Framework for Adaptive Nearest Neighbor Search using Static Features Only\thanks{Hongya Wang is the corresponding author.}
}

\author{
\IEEEauthorblockN{Kaixiang Yang\IEEEauthorrefmark{1} \; Hongya Wang\IEEEauthorrefmark{1}\; Bo Xu\IEEEauthorrefmark{1}; Wei Wang\IEEEauthorrefmark{2}\; Yingyuan Xiao\IEEEauthorrefmark{3}\; Ming Du\IEEEauthorrefmark{1}\; Junfeng Zhou\IEEEauthorrefmark{1}\;}

\IEEEauthorblockA{\IEEEauthorrefmark{1}School of Computer Science and Technology, Donghua University, China}
\IEEEauthorblockA{\IEEEauthorrefmark{2}School of Computer Science and Engineering, University of New South Wales, Australia}
\IEEEauthorblockA{\IEEEauthorrefmark{3}School of Computer Science and Technology, Tianjin University of Technology, Tianjin, China}
\small {ykx@mail.dhu.edu.cn,} \; {\{hywang, xubo, duming, zhoujf\}@dhu.edu.cn,} \; {weiw@cse.unsw.edu.au,} \; {yyxiao@tjut.edu.cn}
}

\maketitle

\begin{abstract}
Approximate nearest neighbor (ANN) search is a fundamental problem in areas such as data management, information retrieval and machine learning. Recently, Li et al. proposed a learned approach named \texttt{AdaptNN} to support adaptive ANN query processing. In the middle of query execution, \texttt{AdaptNN} collects a number of \emph{runtime} features and predicts termination condition for each individual query, by which better end-to-end latency is attained. Despite its efficiency, using runtime features complicates the learning process and leads to performance degradation.


Radically different from \texttt{AdaptNN}, we argue that it is promising to predict termination condition \emph{before} query exetution. Particularly, we developed \texttt{Tao}, a general learning framework for \underline{T}erminating ANN queries \underline{A}daptively using \underline{O}nly static features. Upon the arrival of a query, \texttt{Tao} first maps the query to a \emph{local intrinsic dimension} (LID) number, and then predicts the termination condition using LID instead of runtime features. By decoupling prediction procedure from query execution, \texttt{Tao} eliminates the laborious feature selection process involved in \texttt{AdaptNN}. Besides, two design principles are formulated to guide the application of \texttt{Tao} and improve the explainability of the prediction model. We integrate two state-of-the-art indexing approaches, i.e., IMI and HNSW, into \texttt{Tao}, and evaluate the performance over several million to billion-scale datasets. Experimental results show that, in addition to its simplicity and generality , \texttt{Tao} achieves up to 2.69x speedup even compared to its counterpart, at the same high accuracy targets.
\end{abstract}


\section{Introduction}

Nearest neighbor search is a fundamental problem in domains such as large-scale
image search and information retrieval \cite{DBLP:conf/cikm/LvCL04,
  DBLP:journals/pvldb/LuWWK20}, recommendation\cite{DBLP:conf/www/DasDGR07},
entity resolution\cite{DBLP:conf/cikm/HoffartSNTW12}, and sequence
matching\cite{berlin2015assembling}. Constrained by the curse of dimensionality,
exact nearest neighbor search becomes unaffordable for the rapidly increasing
amount of unstructured data (images, documents and video clips
etc)\cite{DBLP:conf/vldb/WeberSB98}. As a workaround, approximate nearest
neighbor (ANN) search is widely used to provide an appropriate tradeoff between
accuracy and latency.

Currently, the quantization-based and graph-based approaches are two mainstream
ANN search paradigms~\cite{DBLP:journals/pvldb/Qin0X020}. Vector quantization
shrinks database vectors into compact codes based on various quantization
methods, and reduces computation latency and memory requirements by nearly an
order of magnitude~\cite{jegou2010product}. Graph-based
approaches~\cite{malkov2018efficient} first builds a graph to capture proximity
among data points and then performs ANN search by traversing graphs in a greedy
or heuristic fashion. A number of empirical studies show that graph-based
approaches own most appealing tradeoff between accuracy and search cost~\cite{DBLP:journals/corr/abs-2101-12631}.

\begin{figure*}[htbp]
  \centering
  \subcaptionbox{Workflow of \texttt{AdaptNN}}
  {
  \includegraphics[scale=0.20]{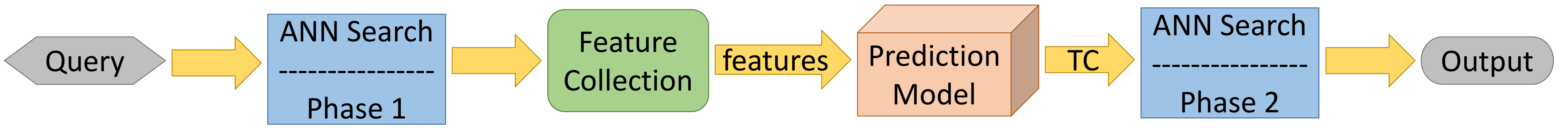}
  \label{fig:pipeline-of-tao:a}
  }

  \subcaptionbox{Workflow of \texttt{Tao}}
  {
    \includegraphics[scale=0.20]{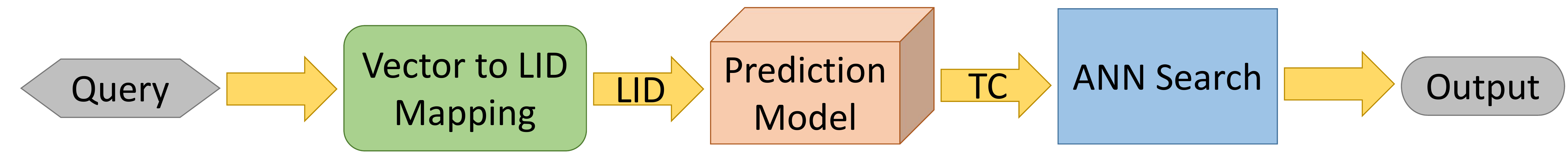}
    \label{fig:pipeline-of-tao:b}
  }

  \caption{Illustrations of \texttt{Tao} and \texttt{AdaptNN}}
  \label{fig:pipeline-of-tao}
\end{figure*}

Both quantization-based and graph-based approaches use fixed configurations that
impose the same termination condition (e.g., the number of candidates to
examine) for all queries, which penalizes ``easy'' queries and incurs
higher-than-necessary running time. 
To address this issue, Li et al.~proposed \texttt{AdaptNN} that predicts and
then applies termination condition for each individual
query~\cite{DBLP:conf/sigmod/LiZAH20}. Figure~\ref{fig:pipeline-of-tao}(a)
illustrates the workflow of \texttt{AdpatNN}. After receiving a query,
\texttt{AdpatNN} first invokes a specific ANN algorithm to execute the query for
a while, then collects a number of runtime
features and predicts the termination condition based on them. Taking the
estimated termination condition as the additional input, the ANN algorithm
continues to run until the condition is met, and outputs the query results
finally.

However, \texttt{AdaptNN} has several limitations despite its efficiency:
\begin{itemize}
\item Firstly, the runtime features require heavy manual
  engineering, that is, these hand-crafted features need to be designed for each algorithms
  individually and no systematic design strategy is available. For example, it
  uses six input features for the HNSW index whereas adopts 14 runtime
  statistics for the IMI index. Hence, it is not easy to adapt \texttt{AdaptNN}
  to other indexing approaches that are not discussed
  in~\cite{DBLP:conf/sigmod/LiZAH20}.
\item Secondly, the end-to-end latency is sensitive to the time interval between
  when \texttt{AdaptNN} makes the prediction (dubbed as \textbf{prediciton time}) and when the query processing
  starts. \cite{DBLP:conf/sigmod/LiZAH20}
  provides no principled method to set the prediction time. As will be
  discussed in detail in Section~\ref{sec:limitations-of-adaptNN}, the optimal
  prediction times depend on specific algorithms and datasets, and the tuning
  procedure is very tedious. In view of this, \texttt{AdaptNN} simply sets it to
  some fixed default value, leading to undesirable performance degradation.
\end{itemize}

Both of the above-mentioned limitations are rooted in the belief that only the
runtime features provide sufficient prediction power to the termination
condition. In this paper, we challenge this belief and propose a simple and
principled \emph{static} feature to perform the prediction accurately.
Specifically, we propose \texttt{Tao}, a general learning framework for
\underline{T}erminating ANN queries \underline{A}daptively using
\underline{O}nly static features. %
Figure~\ref{fig:pipeline-of-tao}(b) illustrates the workflow of \texttt{Tao}.
Upon the arrival of a query, \texttt{Tao} first maps the query to a \emph{local
  intrinsic dimension} (LID) number, and then predicts termination condition
using LID only \emph{before} query execution. The specific ANN search algorithm
takes termination conditions as inputs and answers the query until the condition
is met.

As such, \texttt{Tao} has some advantages worth mentioning:
\begin{itemize}
\item \texttt{Tao} decouples the prediction procedure from query execution and eliminates
  the requirement for tuning the prediction time, making it simple to use
  in practice.
\item \texttt{Tao} employs query vectors and the associated LID, instead of a set of hand-crafted runtime features, as
  model inputs to accomplish the goal of adaptive query processing, making
  it general enough to accommodate new ANN search algorithms

\item \texttt{Tao} improves the explainability of the prediction model by revealing the correlation between LID and search cost and formulating guidelines to use \texttt{Tao}, which is a highly desirable feature in AI-powered applications
  nowadays.
\end{itemize}

To demonstrate the aforementioned advantages, we instantiated two instances of our
framework based on two state-of-the-art indexing approaches
(HNSW~\cite{malkov2018efficient} and IMI~\cite{babenko2014inverted}). Similar
with~\cite{DBLP:conf/sigmod/LiZAH20}, we implement them over the Faiss
similarity search library~\cite{JDH17}. Comprehensive experiments are conducted
to evaluate the end-to-end performance on a collection of real datasets,
including two billion-scale datasets Deep1B~\cite{babenko2016efficient} and
Sift1B~\cite{jegou2011searching}. Empirical results demonstrate that, in addition to its simplicity and generality,
\texttt{Tao} offers up to 2.69x speedup compared with the
runtime prediction strategy.


The contributions of the paper are summarized as follows:
\begin{inparaenum}[(1)]
\item We identify the limitations of the existing approach.
\item We propose and develop a general learning framework for adaptive ANN
  search using only static features.
\item We conduct extensive experiments on various datasets to verify the
  effectiveness and efficiency of our framework.
\end{inparaenum}

\textbf{Roadmap.} Section 2 presents the necessary preliminaries. Section 3 gives a brief review
of \texttt{AdaptNN} and discusses its limitations. Section 4 demonstrates the
feasibility of predicting with static features. Section 5 sketches the workflow
of \texttt{Tao}. Section 6 describes the experimental methodology and reports
the results. Section 7 reviews the related work, and Section 8 concludes the
paper.

\section{Preliminaries}

In this section, we present necessary preliminaries for ANN search and relevant indexing algorithms used in this paper.

\subsection{ANN Search Problem}

Nearest neighbor (NN) search in high dimensional spaces is a challenging problem that has been studied extensively in the past two decades. Formally, let $\mathcal{D} = \{o_1, \cdots, o_n\} \subset \mathbb{R}^d$ be a finite set of $n$ vectors. Given query $q \in \mathbb{R}^d$, $k$NN search returns $k$ results $o_{i}$ $(1 \le i \le k)$, where $o_{i}$ is the $i$-th nearest neighbor of $q$. In this paper we use the Euclidean distance as the distance function to measure (dis)similarity between vectors.

As database sizes reach millions or billions of entries, and the vector dimension grows into the hundreds, any indexing method is shown to be reduced to a linear scan of the whole dataset, which is prohibitively expensive in practice~\cite{jegou2011searching}. Therefore, practitioners often resort to approximate nearest neighbor search. Instead of identifying exact NNs of a query surely, ANN search sometimes may return neighbors that are only close enough to the query. By trading precision for efficiency, ANN search provides better latency than the exact version.

There is a large amount of significant literature on algorithms for approximate nearest neighbor search, which are roughly divided into four categories: tree-structure based approaches, hashing-based approaches, quantization-based approaches and graph-based approaches. In this paper, we focus on quantization-based and graph-based methods due to their appealing performance and popularity in real-life large scale applications~\cite{DBLP:conf/sigmod/WangYGJXLWGLXYY21}.


\subsection{Product Quantization and Inverted Multi-Index}

Product quantization (PQ) algorithm is a lossy compression quantization algorithm that is developed on the basis of vector quantization~\cite{jegou2010product}. By quantizing the vectors of the dataset, it can effectively reduce the storage needed to store the original vectors of the dataset. To the best of our knowledge, quantization-based method is the only one that can support billion-scale datasets on a commodity computer. However, the search is exhaustive for the original PQ -- all database vectors must be evaluated.


To avoid exhaustive search, IVFADC~\cite{jegou2010product} uses two quantization levels to organize large datasets. Particularly, IVFADC groups database vectors into different clusters. When building the index, a list of cluster centroids is trained by $K$-means clustering, and each database vector is assigned to the cluster with the closest centroid. During searching, the index first computes the distances between the query and all cluster centroids, then evaluates database vectors belonging to the first \emph{nprobe} nearest clusters, where \emph{nprobe} is the hyperparameter that should be determined \emph{apriori}.

A number of variants of IVFADC are proposed, among which Inverted Multi-Index (IMI)~\cite{babenko2014inverted} offers the state-of-the-art performance thanks to its efficient space-partitioning strategy. To be specific, IMI decomposes the vectors into several subspaces and trains separate a list of centroids in each subspace, which leads to a fine-grained space partition.

\begin{figure}[htbp]
  \centering
  \includegraphics[scale=0.5]{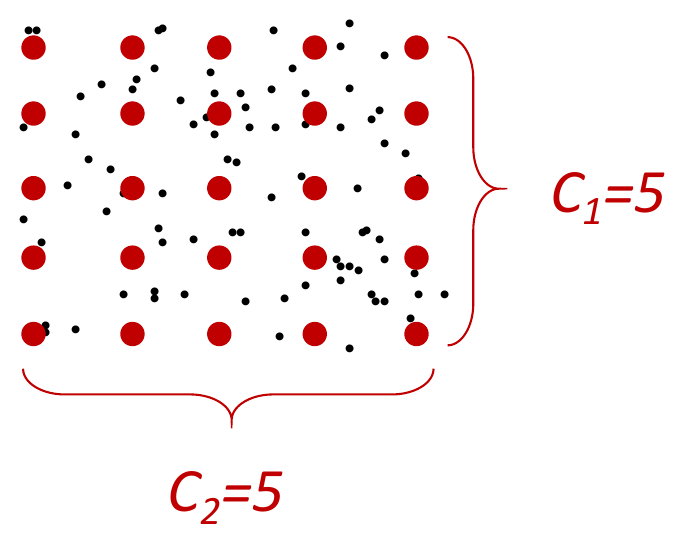}
  \caption{Illustration of IMI (black and red dots represent the training vectors and centroids respectively).}
  \label{fig:diagram-imi}
\end{figure}

For example, Figure~\ref{fig:diagram-imi} illustrates how IMI generates two product codebooks ${C_1}$ and ${C_2}$ for different halves of the vector at the first level. Each codebook contains $5$ sub-codewords, leading to $5^2$ clusters in total. By subtracting a vector to the corresponding cluster centroids, one obtains the residuals for the second quantization level. Similar to IVFADC, IMI searches the same fixed number of \emph{nprobe} nearest clusters for all queries.



\subsection{Graph-based Methods}
Graph-based methods such as hierarchical navigable small worlds (HNSW) are currently one of the most efficient ANN search paradigms \cite{malkov2018efficient}. HNSW employs hierarchical network structures to organized vectors, and such networks are essentially approximate $k$NN-graphs~\cite{DBLP:journals/corr/abs-2012-11083}. To process a query, the graph is traversed starting from the entry point using beam search, and the parameter \emph{efSearch} is used to control the hops of graph traversal. Greater \emph{efSearch} is, more accurate answers will be.

\begin{figure}[htbp]
  \centering
  \includegraphics[scale=0.40]{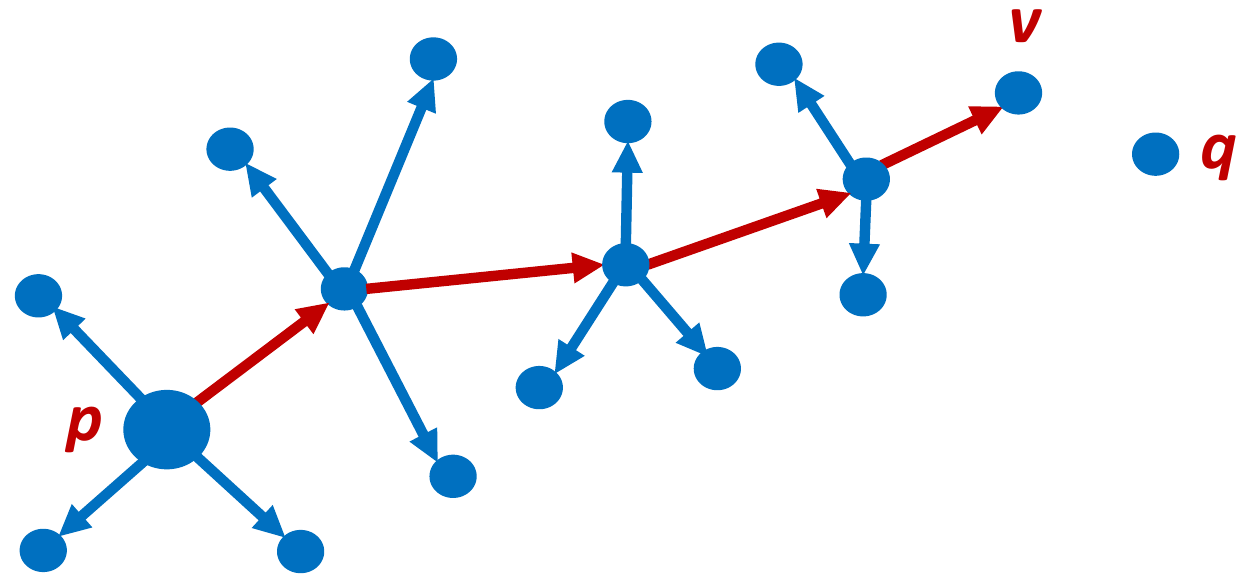}
  \caption{The search procedure of graph-based algorithms.}
  \label{fig:diagram-hnsw}
\end{figure}

For almost all graph-based methods, the ANN search procedure is based on the same principle as illustrated in Figure~\ref{fig:diagram-hnsw}. For a query $q$, start at an initial vertex chosen arbitrarily or using some sophisticated selection rule, say $p$. Moves along an edge to the adjacent vertex with minimum distance to $q$. Repeat this step until the current element $v$ is closer to $q$ than all its neighbors, and then report $v$ as the NN of $q$.


%
%


\section{Motivations}
\subsection{Brief Review of \texttt{AdaptNN}}

Most existing ANN approaches use fixed configurations that apply the same
termination condition to all queries. Li et al.~conducted an empirical study
over three datasets (Deep/Sift/Gist)\footnote{The detailed description of these
  datasets are given in Section~\ref{sect:datasets}} using IMI and HNSW
implementations in the Faiss library \cite{DBLP:conf/sigmod/LiZAH20}. They
observed that, due to the index structures and the vector distributions, the
number of database vectors that must be searched to find the ground-truth
nearest neighbor varies significantly among queries. Take HNSW as an example,
80\% of queries (dubbed ``easy queries'') only need to perform at most
547/481/1260 distance evaluations for Deep10M/Sift10M/Gist, respectively, while
the other 20\% queries (dubbed ``hard queries'') require up to
88696/16618/118277 distance evaluations, respectively.

Fixed configurations lead to high average latency because easy queries are
forced to conduct a lengthy search as the amount required to deliver reasonable
performance for hard queries. To address this issue, Li et al.~propose to
predict different termination condition for each individual query, making easy
queries do less computation than hard ones, thus reducing the end-to-end average
latency dramatically.

One central proposition in \texttt{AdaptNN} is that static features alone (e.g.,
the query vectors themselves) cannot offer good prediction power. To realize
highly accurate predictions, it elaborately identifies a set of \emph{runtime}
features of the intermediate search results, and it adopts gradient boosting
decision trees to predict the minimum amount of search to perform.
As an example, \texttt{AdaptNN} uses the distances between the query and
its 1\emph{st} and 10\emph{th} neighbors as two runtime features for HNSW.

Figure~\ref{fig:pipeline-of-tao}(a) depicts the workflow of \texttt{AdaptNN}.
When receiving a query, \texttt{AdaptNN} first invokes the ANN search algorithm
and performs \emph{a fixed amount of search} to obtain the intermediate search
results. Thereafter, \texttt{AdaptNN} computes the runtime features, coalesces
them with static features, and predicts the termination condition using gradient
boosting decision trees. The termination condition is passed down to the ANN
search algorithm, which continues to run until the condition is met.
Experimental results show that adaptive query processing based on runtime
prediction consistently reduces the average end-to-end latency.

\begin{table*}[htbp]
  \caption{Prediction time vs. Search cost (prediction times are measured in terms of the recall already obtained and the optimal search costs are marked in bold)}
  \begin{center}
  \begin{tabular}{c|c|c|c|c|c|c|c}
  \hline
  \multicolumn{1}{l|}{}     & \multicolumn{7}{c}{\textbf{Search Cost}}                                                                                   \\ \hline
  \textbf{Prediction Time} & \textbf{Deep10M} & \textbf{Sift10M} & \textbf{Gist}  & \textbf{ImageNet} & \textbf{Msong} & \textbf{Trevi} & \textbf{Glove} \\ \hline \hline
  0.5                       & 18,483            & 11,017            & 37,334          & \textbf{71,897}    & 5,337           & 30,621          & \textbf{60,190} \\ \hline
  0.6                       & 22,821            & \textbf{9,453}    & \textbf{34,072} & 86,030             & \textbf{4,521}  & 38,213          & 67,465          \\ \hline
  0.7                       & 18,882            & 10,481            & 34,508          & 87,176             & 8,818           & 22,895          & 66,905          \\ \hline
  0.8                       & 17,594            & 16,787            & 44,390          & 78,641             & 7,182           & \textbf{22,891} & 66,406          \\ \hline
  0.9                       & \textbf{17,033}   & 10,616            & 47,442          & 122,697            & 7,090           & 28,573          & 108,983         \\ \hline
  \end{tabular}
\end{center}
  \label{tab:different_predict_timing}
  \end{table*}

\subsection{Limitations}
\label{sec:limitations-of-adaptNN}
\texttt{AdaptNN}, however, faces a number of limitations. Firstly, the input
features are chosen in an ad-hoc fashion and one has to try different
hand-crafted features for different algorithms. For example, HNSW uses six input
features whereas IMI adopts 14 runtime statistics to predict the minimum amount
of search, respectively. While the importance of different features are
evaluated using the per-feature gain stats from gradient boosting decision tree
models (the importance of a feature is proportional to the total error reduction
contributed by the feature), %
this only alleviate the feature engineering efforts and does not address issues
such as correlation among features. %
Hand-crafted feature engineering impedes the application of
\texttt{AdaptNN} to other ANN search algorithms not addressed
in~\cite{DBLP:conf/sigmod/LiZAH20}.



Secondly, as noted in~\cite{DBLP:conf/sigmod/LiZAH20}, if \texttt{AdaptNN} searches less before the feature generation, the intermediate result features may provide less information gain, reducing the prediction accuracy. If we search more before the feature generation, all queries must search more, increasing the end-to-end average latency. As will be discussed shortly, our preliminary experiments show that the end-to-end latency is quite sensitive to the prediction time. Unfortunately, again, \texttt{AdaptNN} provides no principled method or guideline to determine when to predict in runtime.

The pragmatic (currently practiced) way to choose the prediction time adopted by
\texttt{AdaptNN} is as follows. Given a dataset, manually choose a set of
accuracy targets, say $0.5-0.9$ with a step of $0.1$, and then perform binary
search
to obtain the corresponding search costs (\emph{nprobe} or
\emph{efSearch}) for all accuracy targets. To determine the optimal one, one has
to run \texttt{AdaptNN} using different prediction times over training
datasets
and pick the one with the least average latency. Since the tuning process is
tedious, time-consuming and lossy in nature, \texttt{AdaptNN} simply uses a fixed prediction
time, i.e., the time instant in which 0.8 recall is reached
for all datasets by
default in practice~\cite{DBLP:conf/sigmod/LiZAH20}.

Unsurprisingly, we found that the best prediction time is obviously
data-dependent. Table~\ref{tab:different_predict_timing} lists the total number
of points examined (search cost) under different prediction times (recall
already obtained) for a collection of datasets\footnote{The detailed description
  of these datasets are given in Section~\ref{sect:datasets}} using
\texttt{AdaptNN} with the HNSW method. As one can see 1) the optimal prediction
time vary across different datasets and, 2) choosing inappropriate prediction
time will cause significant performance degradation. For example, the maximum
search cost is twice as much as the minimum for Msong.

The limitations of \texttt{AdaptNN} motivate us to look for a radically
different way to perform adaptive ANN query processing. It is claimed
in~\cite{DBLP:conf/sigmod/LiZAH20} that ``\emph{static features such as the
  query vector itself are not sufficient to predict this termination
  condition}''. Counter-intuitively, however, we observe that query vectors
themselves, with the help of an intermediate feature \emph{local intrinsic
  dimension}, are sufficient to fulfill this goal. Next, we will present the
core ideas and workflow of \texttt{Tao}.

\section{Prediction with Static Features Made Possible}
\label{sec:offline-made-possible}
In this section, we introduce the core notion of our method, i.e., local intrinsic dimension first, and then explore the feasibility of predicting LID using a regression model. The correlation between LID and search cost is examined based on two representative indexing schemes, i.e., IMI and HNSW.



\subsection{Local Intrinsic Dimension}

Many learning tasks involve data represented as vectors of dimension $d$. For example, an image representation is an embedding function that transforms the raw pixel representation of the image to a point in a high-dimensional vector space. While data are embedded in the space $\mathbb{R}^d$, this does not necessarily mean that its intrinsic dimension (ID) is $d$.

The intrinsic dimensionality of a representation refers to the minimum number of parameters (or degrees of freedom) necessary to capture the entire information present in the representation \cite{DBLP:journals/isci/CamastraS16}. Equivalently, it refers to the dimensionality of the $m$-dimensional manifold $\mathcal{M}$ embedded within the $d$-dimensional ambient (representation) space where $m \le d$.



ID has wide applications in many machine learning and data mining contexts. For example, most dimension reduction techniques require that a target dimension be provided by the user. Ideally, the supplied dimension should depend on the intrinsic dimensionality of the data. This has served to motivate the development of models of ID, as well as accurate estimators.

Over the past few decades, many practical models of the intrinsic dimensionality of datasets have been proposed. Examples include the Principal Component Analysis and its variants \cite{DBLP:journals/prl/BouveyronCG11}, as well as several manifold learning techniques \cite{DBLP:journals/nn/KarhunenJ94}. Topological approaches to ID estimate the basis dimension of the tangent space of the data manifold from local samples. Fractal methods such as the Correlation Dimension estimate an intrinsic dimension from the space-filling capacity of the data \cite{DBLP:journals/pami/CamastraV02}. Graph-based methods use the $k$-nearest neighbors graph along with density in order to estimate ID \cite{DBLP:conf/isit/CostaH04}.

The aforementioned intrinsic dimensionality measures can be described as `global', in that they consider the dimensionality of a given set as a whole, without any individual object being given a special role. In other words, all vectors share the same intrinsic dimension. In contrast, `local' ID measures are defined as those that involve only the $k$-nearest neighbor distances of a specific location in the space.

Several local intrinsic dimensionality models have been proposed recently, such as the expansion dimension \cite{DBLP:conf/stoc/KargerR02}, the generalized expansion dimension \cite{DBLP:conf/icdm/HouleKN12}, the minimum neighbor distance \cite{DBLP:journals/ml/RozzaLCCC12}, and local continuous intrinsic dimension (LID)\cite{DBLP:conf/icdm/Houle13}. In this paper we focus on LID, which is defined formally as follows:


\begin{myDef}
	\label{def1}
    (~\cite{DBLP:conf/icdm/Houle13}.) Given an absolutely continuous random distance variable $X$, for any distance threshold $x$ such that the cumulative density function $F_X(x) > 0$, the local continuous intrinsic dimension of $X$ at distance $x$ is given by
    $$
    LID_X\left( x \right) \overset{\bigtriangleup}{=}\lim _{\epsilon \rightarrow 0^+}\frac{\ln F_X\left( \left( 1+\epsilon \right) x-\ln F_X\left( x \right) \right)}{\ln \left( 1+\epsilon \right)}
    $$
    wherever the limit exists.
\end{myDef}

LID quantifies intrinsic dimension in terms of the rate at which the number of encountered objects grows as the considered range of distances expands from a reference location. Estimates of local ID constitute a measure of the complexity of data and LID could give researchers and practitioners more insight into the nature of their data, and therefore help them improve the efficiency and efficacy of their applications \cite{DBLP:journals/tods/ShaftR06}.


Amsaleg et al. studied several estimators of LID using maximum likelihood estimation (MLE), the method of moment, probability weighted moment and regularly varying functions~\cite{DBLP:conf/kdd/AmsalegCFGHKN15}. Experimental results show that the performance of different estimators to be largely in agreement with one another, and faster initial convergence favors the choice of MLE for applications where the number of available query-to-neighbor distances is limited, or where time complexity is an issue.

Assume that we are given a sequence $x_1$ ,..., $x_n$ of observations of a random distance variable $X$ with support [0,$w$) in ascending order, that is, $x_1$ $\le$ $x_2$ $\le$ ··· $\le$ $x_n$. Then, the MLE estimator of LID can be calculated as

\begin{equation}
  \label{equ:lid-MLE-estimation}
  \begin{aligned}
    \widehat{LID}_X \overset{\bigtriangleup}{=} \left( \frac{1}{k}\sum_{i=1}^k{\ln \left( \frac{w}{x_i} \right)} \right) ^{-1}
  \end{aligned}
  \end{equation}




\begin{figure}[htbp]
  \centering
  \subcaptionbox{Deep10M}
  {
  \includegraphics[width = .46\linewidth]{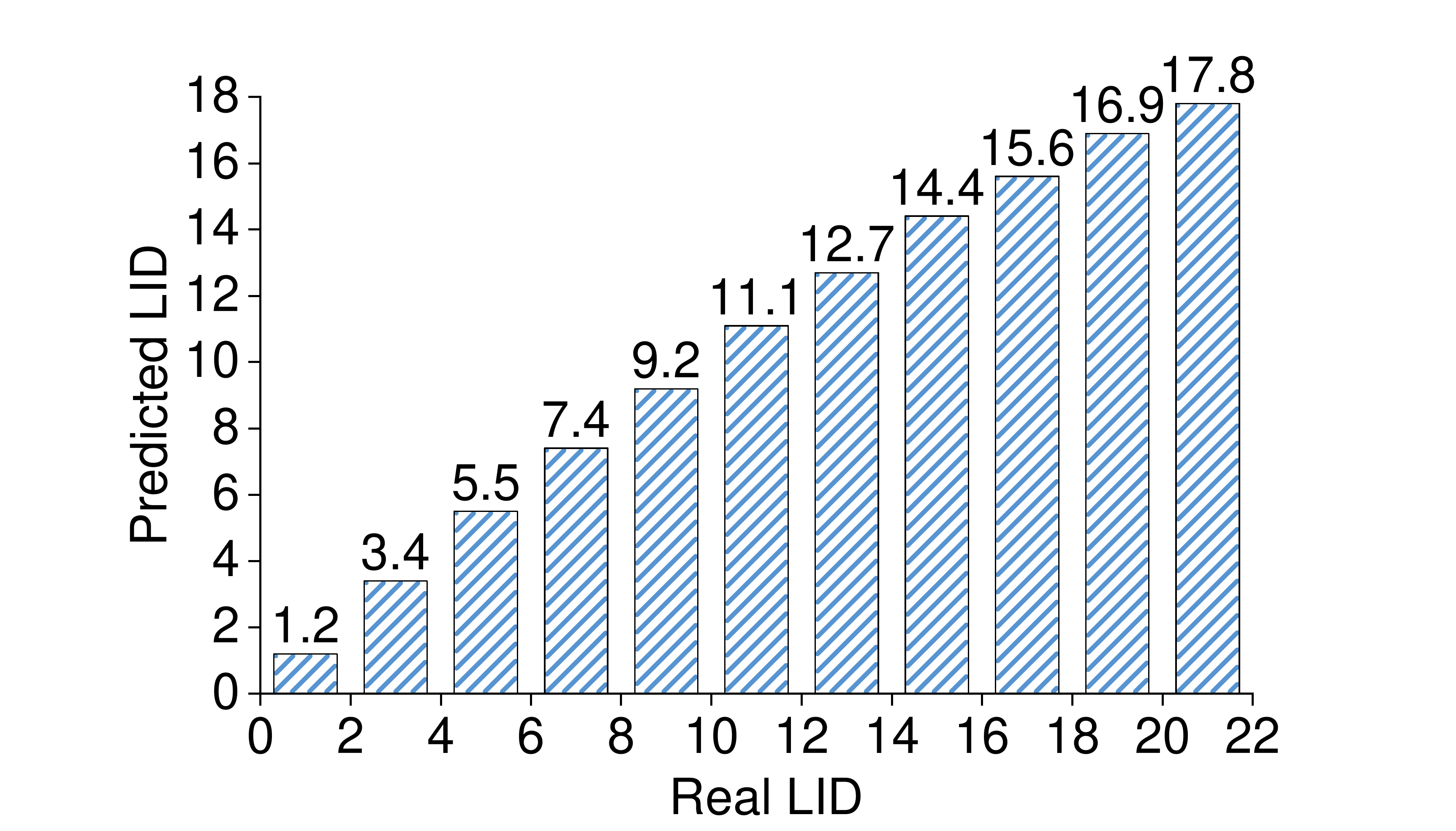}
  \label{fig:real-and-predicted-lid-sift:a}
  }
  \subcaptionbox{Sift10M}
  {
    \includegraphics[width = .46\linewidth]{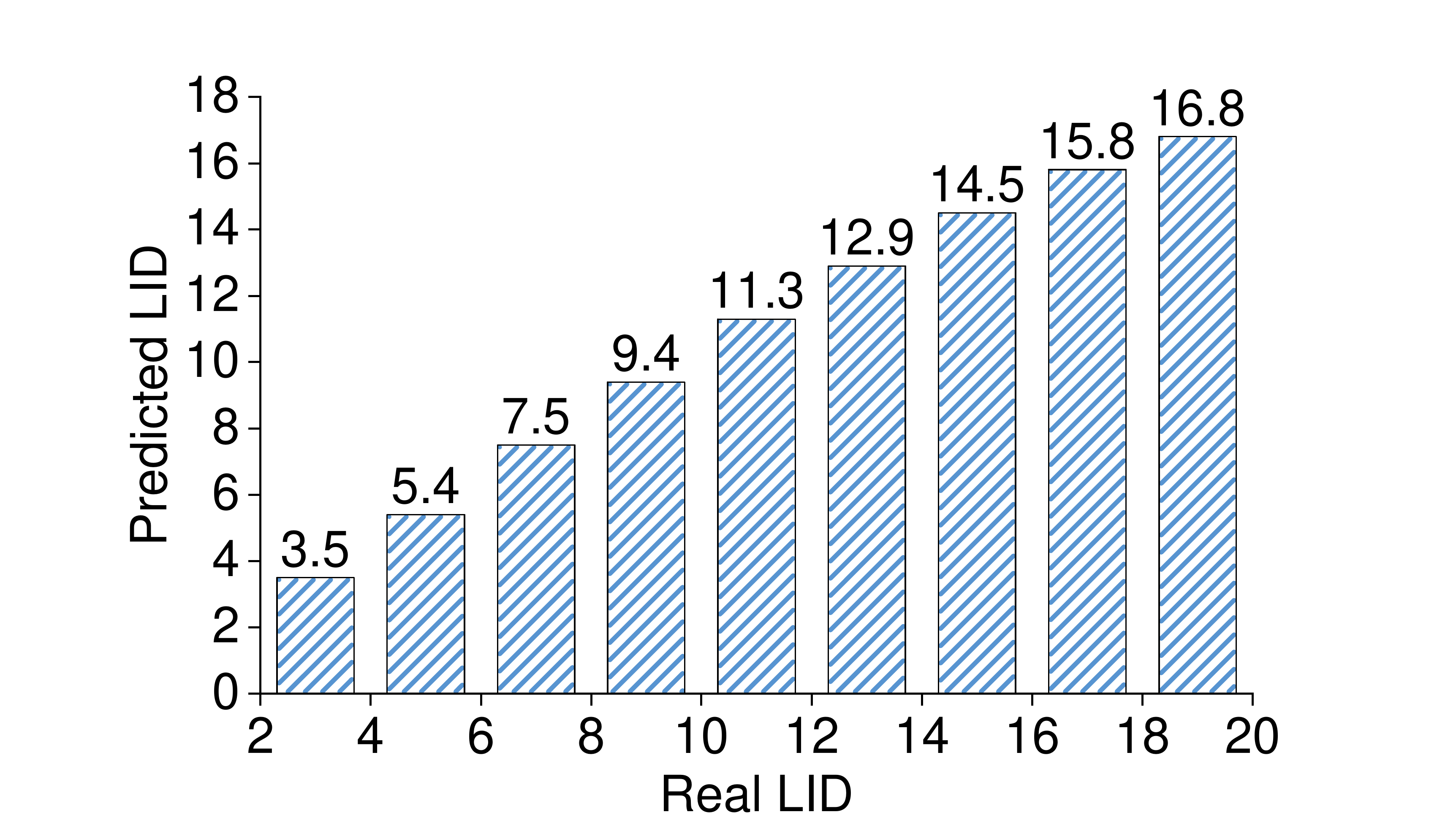}
    \label{fig:real-and-predicted-lid-sift:b}
  }

  \caption{Real vs. Predicted LID on Deep10M and Sift10M.}
  \label{fig:real-and-predicted-lid-sift}
\end{figure}


\begin{table}[htbp]
  \caption{MAE, RMSE and R2Score for different datasets.}
  \begin{center}
  \begin{tabular}{c|c|c|c}
  \hline
  & \textbf{MAE} & \textbf{RMSE} & \textbf{R2Score} \\
  \hline
  \hline
Deep10M   & 1.15  & 1.88  & 0.68  \\ \hline
Sift10M   & 0.65  & 0.89  & 0.89  \\ \hline
Gist   & 1.24  & 1.72  & 0.89  \\ \hline
ImageNet   & 0.77  & 1.07  & 0.91  \\ \hline
Glove   & 1.80  & 2.48  & 0.88  \\ \hline
MSong   & 0.67  & 1.22  & 0.80  \\ \hline
Trevi   & 3.17  & 4.41  & 0.82  \\ \hline
\end{tabular}
\label{tab:mae}
\end{center}
\end{table}

\subsection{Vector to LID Mapping}

By definition, the LID number of a vector $q$ is closed related to the distribution of its near neighbors, which suggests that LID might be a promising indicator for the difficulty in finding $k$NN of $q$. Unfortunately, for an unsee query, there is no way to calculate the LID value without knowing its $k$NN. To circumvent this dilemma, we explore the possibility of predicting LID for unseen vectors using neural networks, considering they are extremely good at capturing complex relationships in high dimensional spaces~\cite{DBLP:conf/sigmod/KraskaBCDP18}.

Suppose dataset $\mathcal{D}$ is a sample drawn from a population that follows some unknown probability distribution $\mathcal{X}$. For any point $o \sim \mathcal{X}$, we can calculate the LID estimate of $o$ using Equation~\eqref{equ:lid-MLE-estimation}, where $x_i$ is the distance between $o$ and its $i$-th NN in $\mathcal{D}$. For unseen vectors, assume there exists a function $f(\cdot)$ that relates any sample vector drawn from $\mathcal{X}$ to a single LID value. The \emph{universal approximation theorem} tells us that a neural network can approximate $f(\cdot)$, given that $f(\cdot)$ is continuous on a compact set and the neural network has sufficiently many hidden neurons with activation functions \cite{DBLP:journals/mcss/Cybenko89, DBLP:journals/nn/Hornik91}.

Verifying the continuity of $f(\cdot)$ analytically is impossible because we do not even have an closed-form expression for $f(\cdot)$. To this end, we demonstrate the feasibility by empirically evaluating the approximation errors on several real datasets. Particularly, we trained an MLP network with two hidden fully-connected layers, using 200 neurons per layer (i.e., 200 width) and ReLU activation functions. The inputs to the neural network are exclusively the original vectors and the outputs are the predicted LID values.  We use the training vectors listed in Table~\ref{tab:dataset} to generate training data and the query vectors to generate testing data.



Table~\ref{tab:mae} summarizes the statistics of common evaluation metrics for regression models. For mean absolute error (MAE) and RMSE, lower is better. For R2 scores, higher is better. As we can see from the table, we can reliably estimate the LID of query vectors using our regression model. E.g., the MAE of the prediction is less than 1.24 in 6 out of the 8 datasets. We formulate this observation as Property 1:

\begin{property}
\label{pro:vectors-to-lid}
Given a dataset $\mathcal{D}$, the LID values of high-dimensional vectors in $\mathcal{D}$ can be estimated using a practical regression model with small approximation error.
\end{property}

We further illustrate the real and the predicted LID values in
Figure~\ref{fig:real-and-predicted-lid-sift} over Deep10M and Sift10M. The $x$-axis represents the LID numbers of vectors calculated using Equation~\eqref{equ:lid-MLE-estimation}, which are partitioned into bins of size 2. The average of predicted LID values in each bin are marked on the top of bars. As we can see, the estimates match rather well with the real values for small LID ($0-14$ for Deep10M and $2-16$ for Sift10M). For large LID, the
regression model somewhat underestimates the target values, mainly due to the
sparsity of training vectors with large LID numbers. Please note that similar trends are observed for other datasets and we do not report them here due to space limitation.

We also adopt gradient boosting decision tree models (using the LightGBM library~\cite{DBLP:conf/nips/KeMFWCMYL17}) to perform the vector to LID mapping, which achieves similar results with MLP. This suggests that there exists inherent correlation between vectors and LID for all real datasets that we have experimented with, and such correlation may be model independent to some extent. We conjecture that the reason might be these datasets, while represented in high-dimensional spaces, inherently resides on some low-dimensional manifolds \cite{2008Neural, DBLP:journals/tnn/Andras14}. As a result, the relation between vectors and LID can be easily captured using a reasonable regression model. We believe that more in-depth analysis of this correlation deserves further study on its own right, and is thus out of the scope of this paper.


\begin{figure}[htbp]
  \centering
  \subcaptionbox{Deep10M}
  {
  \includegraphics[width = .46\linewidth]{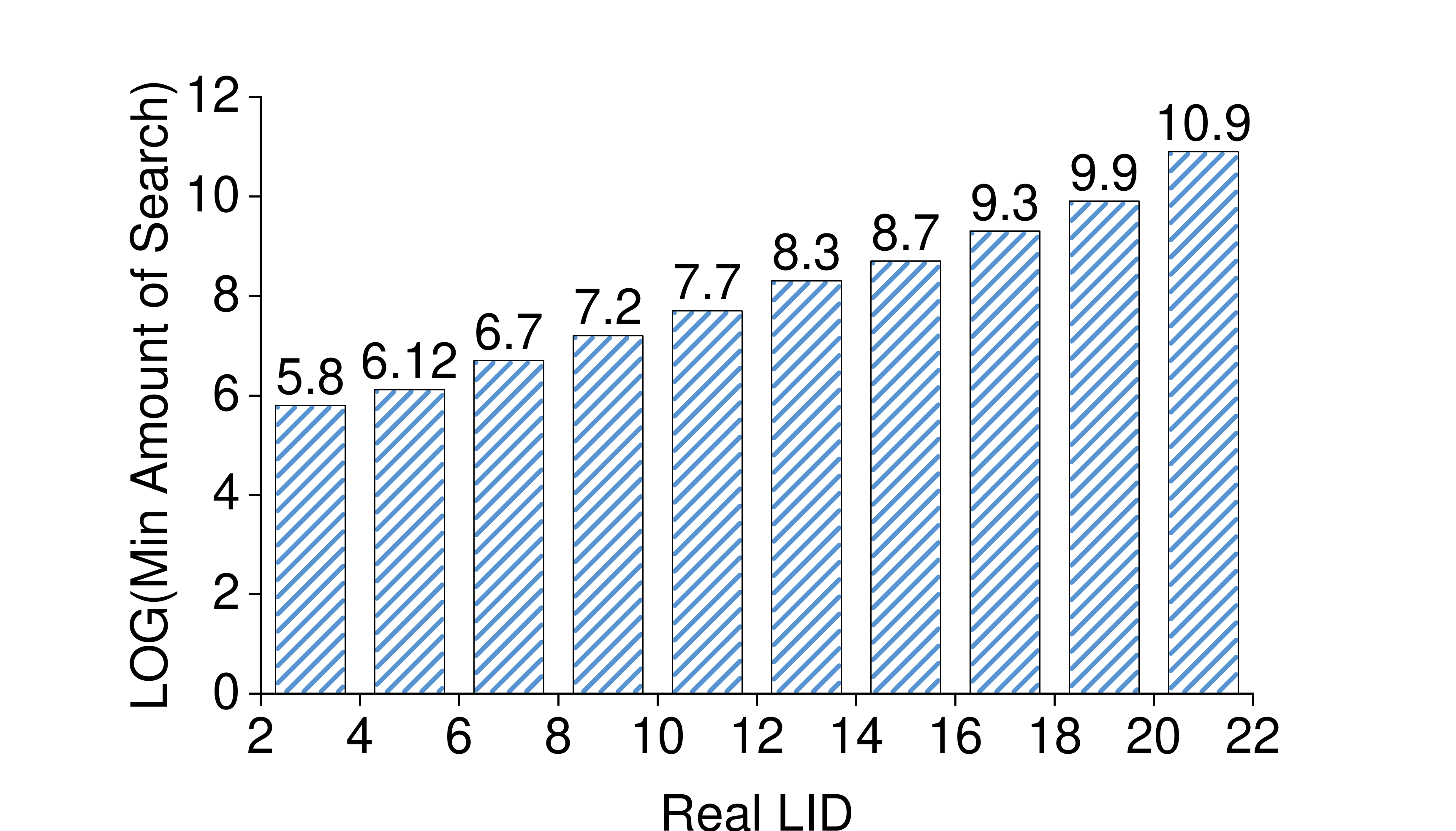}
  \label{fig:real_lid_and_term_cond}
  }
  \subcaptionbox{Sift10M}
  {
    \includegraphics[width = .46\linewidth]{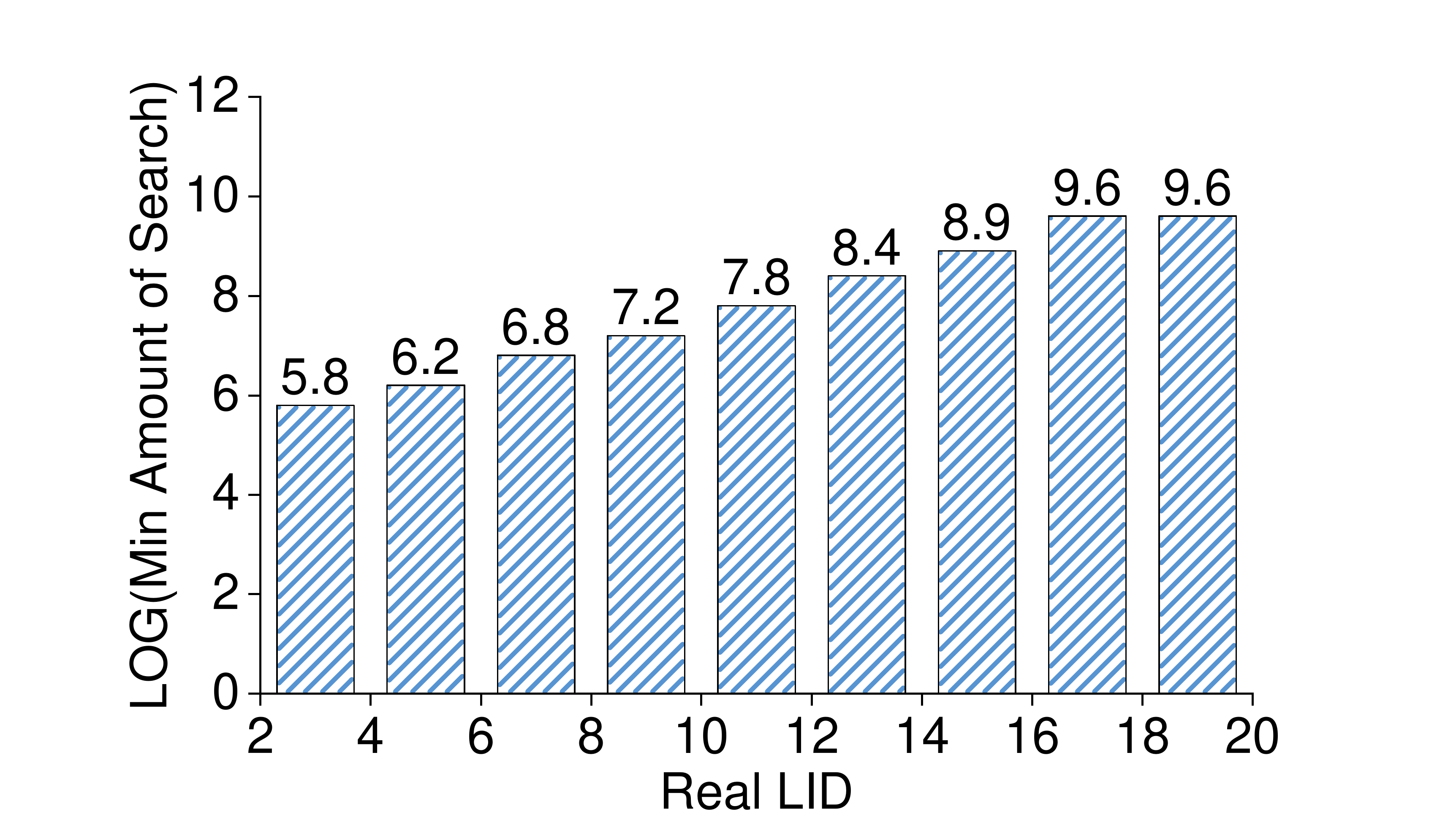}
    \label{fig:real_lid_and_term_cond_sift}
  }

  \caption{Search cost vs. LID for HNSW.}
  \label{fig:real_lid_and_term_cond_all}
\end{figure}

\begin{figure}[htbp]
  \centering
  \subcaptionbox{Deep1B}
  {
  \includegraphics[width = .46\linewidth]{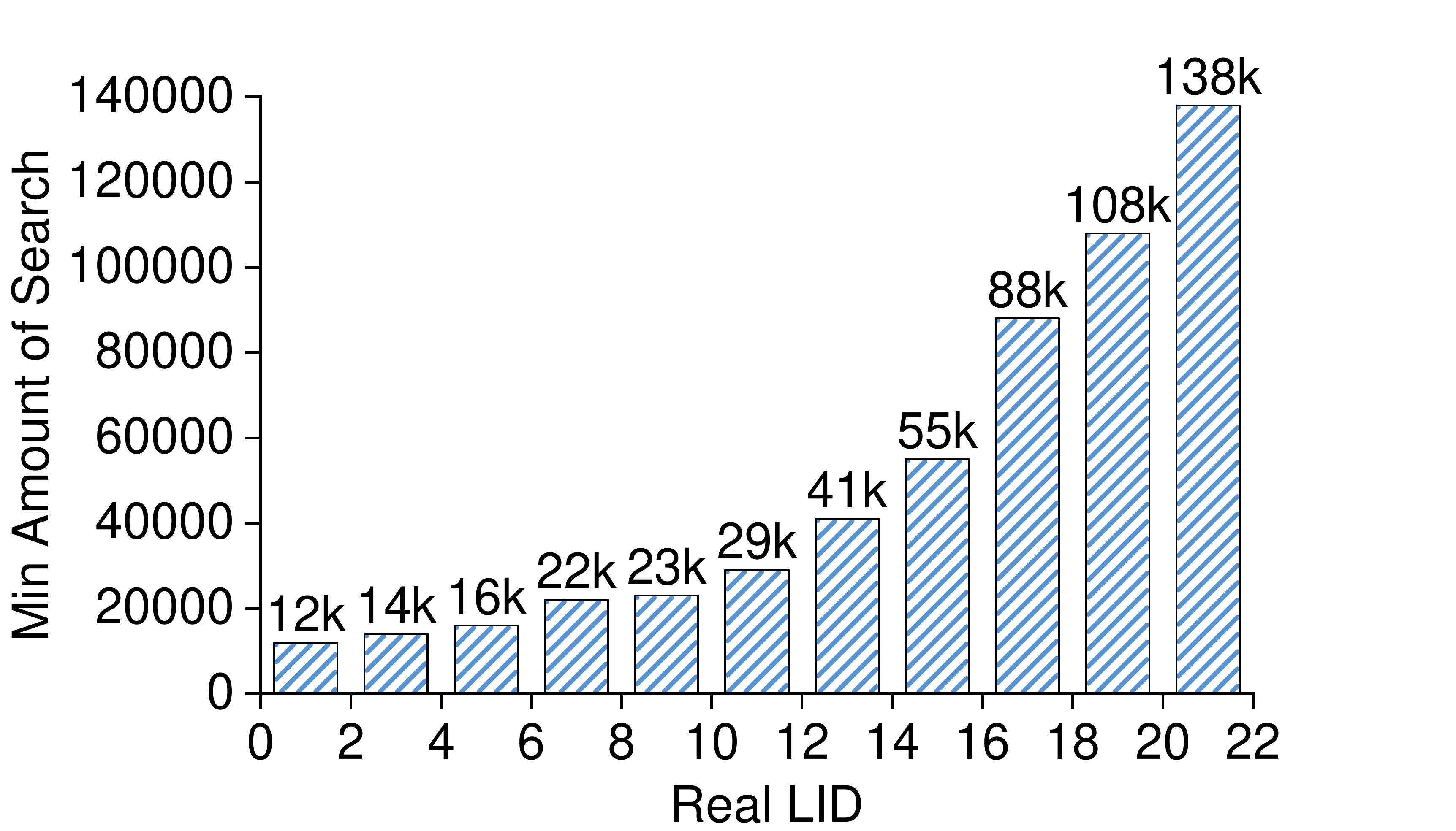}
  \label{fig:real_lid_and_term_cond_imi}
  }
  \subcaptionbox{Sift1B}
  {
    \includegraphics[width = .46\linewidth]{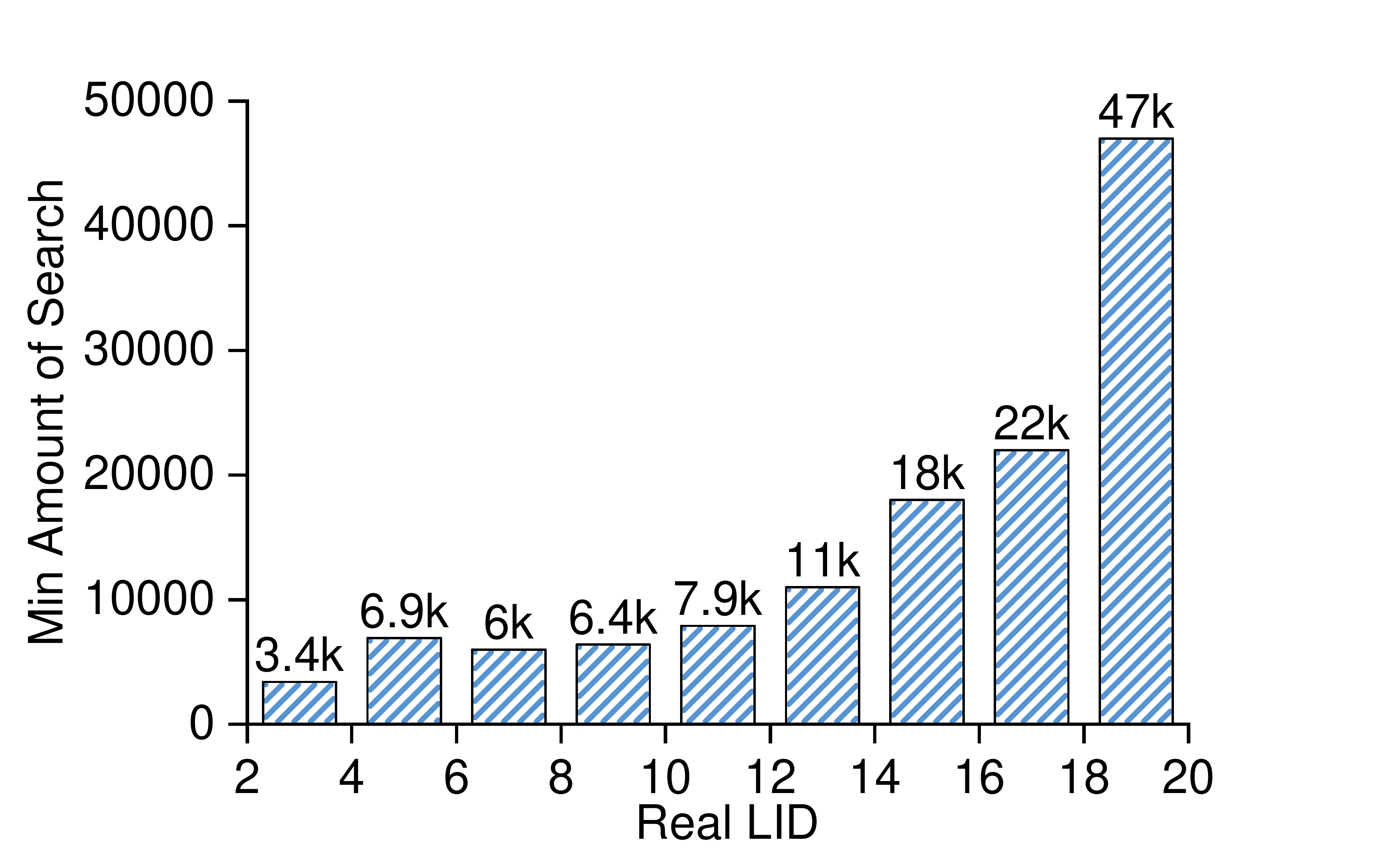}
    \label{fig:real_lid_and_term_cond_imi_sift}
  }

  \caption{Search cost vs. LID for IMI.}
  \label{fig:real-and-predicted-lid-sift-all}
\end{figure}

\subsection{Correlation between LID and Search Cost}

LID is aimed to quantify the local intrinsic dimensionality of a feature space exclusively in terms of the distance distribution of neighbors of a specific location. It is well known that the dimensionality of the space that the point resides in has a significant impact on the difficulty of NN search~\cite{DBLP:conf/vldb/GionisIM99}. Intuitively, the overhead of NN search is probably correlated with the LID of query vectors.


The way to measure search cost varies over different indexing methods. For IMI, the amount of search is represented by the number of nearest clusters that need to be examined (\emph{nprobe}). For HNSW, we use the number of distance evaluations to represent the amount of search. This is because 1) The distance evaluation between query and database vector is a time-consuming task; 2) The number of distance evaluations fluctuates greatly even with the same number of hops in the graph (\emph{efsearch}).




Figure~\ref{fig:real_lid_and_term_cond_all} illustrates the mean minimum search cost to find the ground truth nearest neighbor for query vectors with different LID using HNSW. The bin size is 2 and the search cost is averaged over each bin. A base 2 logarithmic scale is used for $y$-axis. As one can see, for both Deep10M and Sift10M, the search cost increases roughly exponentially as LID grows, suggesting that LID is a promising static feature to predict the termination condition.





Figure~\ref{fig:real-and-predicted-lid-sift-all} shows the histogram of the mean minimum number of centroids examined to find the ground truth nearest neighbor using IMI. The LID bin size is set to 2 and the search cost is averaged over each bin. As with HNSW, there exists a strong correlation between LID and the number of centroids examined. It is worth mentioning that IMI and HNSW exhibit similar trends for the remaining datasets that have been experimented with, which are not reported here due to space limitation. We highlight such correlations as Property 2:

\begin{property}
\label{pro:lid-to-search-cost}
Given an ANN search algorithm $\mathcal{A}$ and a dataset $\mathcal{D}$, the LID values of query vectors are positively correlated with the minimum amount of search to identify the true nearest neighbors in $\mathcal{D}$ by $\mathcal{A}$.
\end{property}

\subsection{Remarks}
As discussed above, Property 1 is data dependent and Property 2 depends on both data and algorithms. Given an ANN search algorithm and a dataset, it is possible to accomplish the goal of adaptive query processing as long as these two properties hold. Such simple guidelines turn the learning black box transparent to users to some extent, eliminate the hand-crafted feature selection process and ease the applications of $\texttt{Tao}$ to other ANN algorithms and datasets.




\section{Adaptive ANN Search with \texttt{Tao}}

In this section, we describe the prediction pipeline of the proposed learning framework and how ANN search algorithms, exemplified by IMI and HNSW, are integrated into \texttt{Tao}.

\textbf{The general workflow.} Our prediction pipeline consists of two phases as illustrated in Figure~\ref{fig:pipeline-of-tao}. In Phase 1, the regression model takes query vectors as inputs and outputs predicted LID numbers. In Phase 2, the other regression model accepts a LID value and reports a numerical indicator, suggesting how much search should be done for the query.


\textbf{The inputs.} Instead of combining a number of hand-crafted static and runtime features as inputs, \texttt{Tao} needs only a single independent variable, that is, the query vector itself.

\textbf{The output.} For each query, we expect to predict the minimum amount of search to obtain the ground truth nearest neighbor. Different indexing approaches may have different metrics to quantify the search cost, but what we need is often a numerical value which is proportional to the search latency.


\textbf{Model Selection and Training.}
According to the two properties we formulated, \texttt{Tao} is actually  model independent, meaning that any reasonable regression model can be used to fulfill the vector$\rightarrow$LID$\rightarrow$termination-condition prediction framework. In this paper, we choose two popular models, i.e., MLP and gradient boosting decision trees, to show the effectiveness of \texttt{Tao}. For MLP, we employ two distinct neural networks to fulfill the predictive tasks in different phases, where the standard feed forward structure with two fully-connected hidden layers is adopted and the ReLU activation function is used across all layers. The parameters of these two neural networks will be discussed in detail in Section~\ref{sec:benchmark-methods}.

For the second one, we build and train two distinct gradient boosting decision tree models using the LightGBM library~\cite{DBLP:conf/nips/KeMFWCMYL17} to accomplish the predictive tasks needed by \texttt{Tao}. Since the prediction performance of LightGBM is similar to that of MLP, we only report the experimental results using MLP in this paper due to space limitation.

We use the training vectors in Table~\ref{tab:dataset} to generate training/validation data and use the query vectors to generate testing data. Each vector generates one row of data which includes both the output target value and the true LID value. To obtain output target values, we need to first perform an exhaustive search to find the ground truth nearest neighbor(s), and then find the minimum amount of search to reach (one of) it. To calculate LID, we identify the top-1000 neighbors of each vector in the training set and compute LID values using Equation~\eqref{equ:lid-MLE-estimation}. We trained the two neural networks independently, i.e., no joint learning techniques are used. For the first one, we choose a fixed number of training epochs (200) and batch sizes varying from 200 to 1000, depending on the sizes of training sets. For the second one, a smaller number of training epochs is used (20) and batch sizes are the same as the first one.

\textbf{Integration with IMI and HNSW.} The integration of \texttt{Tao} with the existing indexing algorithms such as IMI and HNSW is noninvasive, in that very few modifications have to be done with them. Particularly, since \texttt{Tao} decouples the prediction models from query execution, we only need to change the termination conditions in the code bases, involving less than 5 lines of the core codes for IMI and HNSW.

For IMI, the number of nearest clusters to search equals \emph{Max}(\emph{thresh}, \emph{multiplier}*\emph{$2^{TC}$}), where the \emph{thresh} equals the maximum target value in the training data and $TC$ is the predicted search cost. For HNSW,
the number of distance evaluations equals \emph{Max}(\emph{thresh}, \emph{multiplier}*\emph{$2^{TC}$}). Similar to~\cite{DBLP:conf/sigmod/LiZAH20}, \emph{multiplier} is the scale factor needs to be tuned in order to achieve given accuracy target.


Algorithm 1 sketches how \texttt{Tao} is integrated with HNSW.

\begin{algorithm}
  \caption{\texttt{Tao} with HNSW}
  \label{alg:HNSW}
  \begin{algorithmic}

  \renewcommand{\algorithmicrequire}{\textbf{Input:}}
  \renewcommand{\algorithmicensure}{\textbf{Output:}}
  \REQUIRE Query vector: $q$,\\
    Entry point: $ep$,\\
    Number of nearest neighbors to return: $k$,\\
    Candidate queue: $cq$, \\
    Maxheap list: $w$

    \ENSURE $k$ nearest neighbors to q

    \STATE $tc$ $\gets$ predict with \texttt{Tao}
    \STATE $ndis \gets 0$ //number of distance comparisons\\
    \STATE {$cq.push\_back(ep)$}

    \WHILE {$|cq| > 0$}
    \STATE {$v = cq.top()$}
    \STATE {$cq.pop()$}

    \FOR { $v.neighbors$}
      \STATE calcultate distances and update $cq,w$
      \STATE {$ndis++$}
    \ENDFOR

    \IF {$ndis > tc$}
      \STATE {\textbf{break}}
    \ENDIF
  \ENDWHILE
  \RETURN $k$ nearest neighbors to $q$ in $w$
  \end{algorithmic}
\end{algorithm}






\section{Evaluation}
In this section, we evaluate the performance of \texttt{Tao} by implementing it in the Faiss ANN search library (CPU version) \cite{DBLP:conf/sigmod/LiZAH20} similar to \texttt{AdaptNN}, which makes the comparison fair and reasonable. All experiments are carried out on a server with E5-2620 v4@2.10GHz CPU and 256GB memory.


\subsection{Datasets}
\label{sect:datasets}
We used seven publicly available real datasets, which are of different data dimensions, cardinalities and types. Table~\ref{tab:dataset} lists the statistics of datasets we have experimented with. The sizes of these datasets range from millions to one billion.

\begin{table}[htbp]
\caption{Dataset Summary}
\begin{center}
\begin{tabular}{c|c|c|c|c|c}
\hline
\textbf{Dataset} & \textbf{Dim.} & \textbf{Base}  & \textbf{Training} & \textbf{Query} & \textbf{Type} \\ \hline \hline
Deep  & 96    & 10M,1B & 1M    & 10,000 & Image \\ \hline
Sift  & 128   & 10M,1B & 1M    & 10,000 & Image \\ \hline
Gist  & 960   & 1M    & 0.5M  & 1,000 & Image \\ \hline
ImageNet  & 150   & 2.34M & 200,000 & 10,000 & Image \\ \hline
Msong & 420   & 9.22M & 100,000 & 10,000 & Audio \\ \hline
Trevi & 4096  & 1M    & 20,000 & 1,000 & Image \\ \hline
Glove & 100   & 1.19M & 100,000 & 10,000 & Text \\ \hline
\end{tabular}
\label{tab:dataset}
\end{center}
\end{table}

\begin{itemize}
\item \textbf{Deep1B}\footnote{https://github.com/facebookresearch/faiss/tree/master/} consist of 1 billion contains deep neural codes of natural images obtained from the activations of a convolutional neural network.

\item \textbf{Deep10M}\footnote{https://github.com/facebookresearch/faiss/tree/master/} is a subset of Deep1B.


\item \textbf{Sift1B}\footnote{http://corpus-texmex.irisa.fr/} consist of 1 billion 128-dim SIFT vectors.

\item  \textbf{Sift10M}\footnote{https://archive.ics.uci.edu/ml/datasets/SIFT10M} consists of 10 million 128-dim SIFT vectors.

\item \textbf{Gist}\footnote{http://corpus-texmex.irisa.fr/} is an image dataset which contains about 1 million data points.

\item \textbf{ImageNet}\footnote{ http://cloudcv.org/objdetect/} contains about 2.4 million data points with 150 dimensions dense SIFT features.

\item \textbf{Msong}\footnote{http://www.ifs.tuwien.ac.at/mir/msd/download.html} is a collection of audio features and metadata for millions of contemporary popular music tracks.

\item \textbf{Trevi}\footnote{ http://phototour.cs.washington.edu/patches/default.htm} consists of around 1,000,000 4096-$d$ vectors.

\item \textbf{Glove}\footnote{ http://nlp.stanford.edu/projects/glove/} contains 1.2 million 100-$d$ word feature vectors extracted from Tweets.

\end{itemize}

For ImageNet/Msong/Trevi/Glove datasets, due to these datasets do not have a training set or the query set is small, so we split a subset of the database as the training set or query set. Taking ImageNet as an example, we split the first 200,000 of the database into a training set, then we split 200,001 to 210,000 of the database into a query set, and the rest as the new ImageNet base set.

\subsection{Benchmark Methods}
\label{sec:benchmark-methods}

\begin{itemize}
\item \textbf{Fixed Configuration.} The implementations of IMI and HNSW in the Faiss ANN search library (CPU version) \cite{DBLP:conf/sigmod/LiZAH20} is used. Parameters to control the answer quality, that is, \emph{efSearch} (HNSW) and \emph{nprobe} (IMI), are tuned to achieve the given target accuracy, and fixed for all queries.

\item \textbf{\texttt{AdaptNN}.} We use the training and parameter tuning methods described in \cite{DBLP:conf/sigmod/LiZAH20} to obtain the \emph{optimal} prediction time, which is both algorithm and dataset dependent. In other words, we report the \emph{best} end-to-end search performance that \texttt{AdaptNN} can deliver for any given algorithm and dataset.

\item \textbf{\texttt{Tao}.} \texttt{Tao} employs two neural networks as the regression models. Both neural networks employ two fully connected layers with ReLU activation functions. For the first one, 200 neurons are used for each fully connected layer. The second neural network uses 10 neurons for each fully connected layer.

\item \textbf{Vector Only (VO).}  VO combines the two neural networks of \texttt{Tao} into one single MLP with five hidden layers. To make fair comparison, the structure of VO is the same as those of \texttt{Tao}, i.e., five hidden layers are fully connected and the number of neurons in each layer are set to 200, 200, 1, 10, 10 respectively. The only difference between VO and \texttt{Tao} is the training process -- LID is not used explicitly in VO and the input and output of the neural network of VO are query vectors and the amount of search to find the true NN, respectively.


\end{itemize}

\subsection{Performance Metrics}

The end-to-end latency and accuracy are two important metrics for evaluating ANN algorithms. To compare the performance of the baselines and proposed approach, we perform controlled experiments to keep the accuracy achieved by all approaches at the same level, and then to compare the average latency. Given an accuracy target, we perform binary search to find the minimum average latency for the fixed configuration baseline. For \texttt{AdaptNN} and \texttt{Tao}, we multiply the predicted search cost with a coefficient (tuning knob) to reach this desired accuracy. Then we compare their performance at each accuracy target, where the prediction overhead is included in the end-to-end latency.

For the accuracy target, we use recall$@$1 (the fraction of queries where the top-1 nearest neighbor returned from search is (one of) the ground truth nearest neighbor) for HNSW. For IMI, we use recall$@$100 (the fraction of queries where the top-100 nearest neighbors returned from search include (one of) the ground truth nearest neighbor) as the accuracy target since it’s challenging for quantization-based approaches to reach high recall$@$1. We measure the average latency in the single-threaded setting as in previous work~\cite{DBLP:conf/sigmod/LiZAH20}.

\subsection{Prediction Overhead}

Since the parameters of neural networks are fixed for all datasets, the model size is a constant (560KB) in our experiment setting as well, which is similar to those of \texttt{AdaptNN} (274 KB to 310 KB). When making prediction, it takes around 65us using the Keras framework for \texttt{Tao} (7us to 47us for \texttt{AdaptNN}). While the prediction overhead is very small already, a significant reduction to a few nano seconds is possible if one uses a plain C++ implementation~\cite{DBLP:conf/sigmod/KraskaBCDP18}. We leave this as future work since it is not a dominant factor in the end-to-end latency.

\begin{figure}
  \centering
  \includegraphics[scale=0.22]{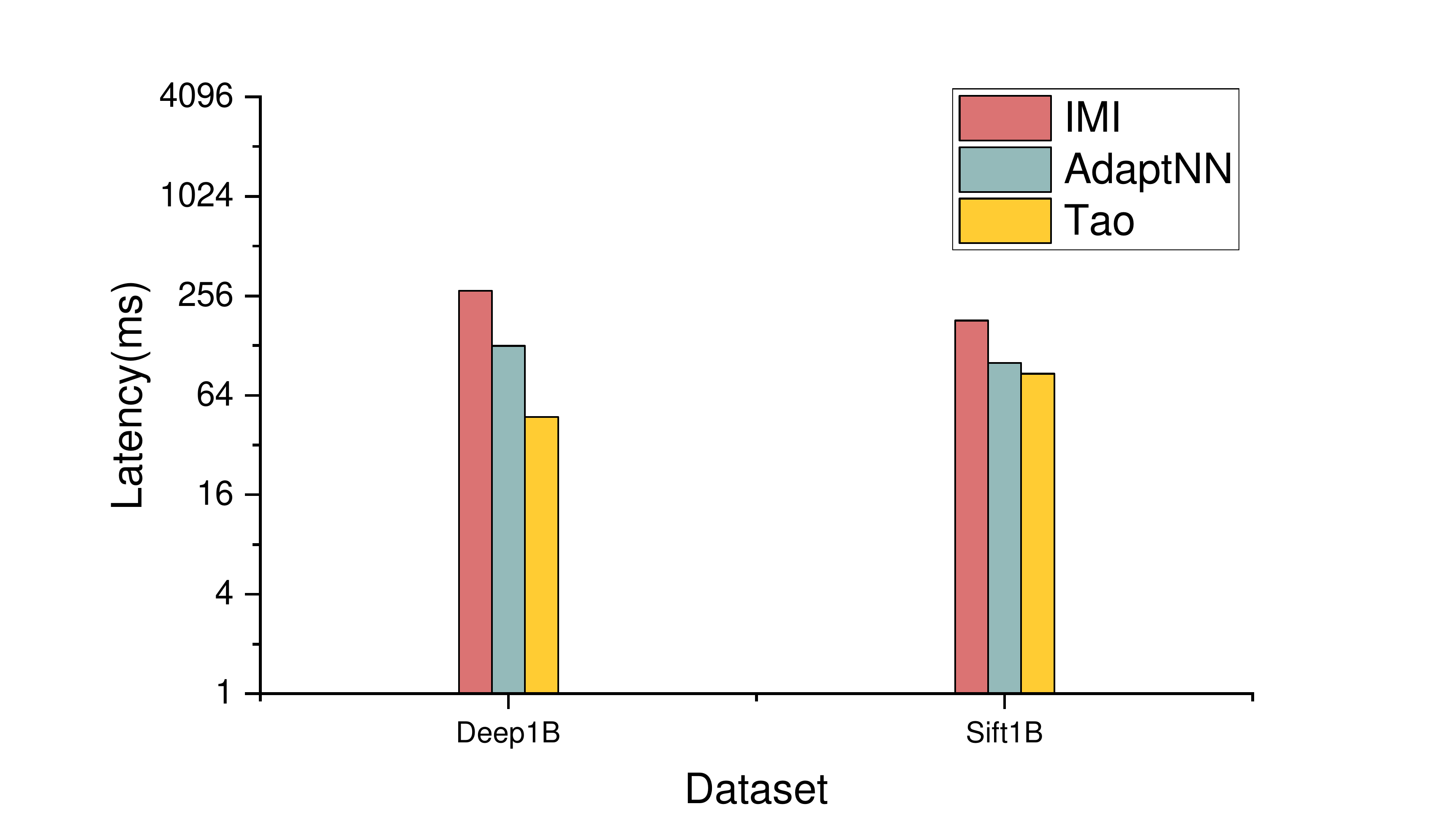}
  \caption{Latency for plain IMI, \texttt{AdaptNN} and \texttt{Tao}.}
  \label{fig:latency_imi}
\end{figure}

\begin{table*}[htbp]
  \caption{Latency for IMI, \texttt{AdpatNN} and \texttt{Tao}.}
  \begin{center}
  \begin{tabular}{c|c|c|c|c|c|c}
  \hline
  \textbf{Dataset} &\textbf{Recall} & \textbf{IMI} & \texttt{AdpatNN} & \texttt{Tao} & \textbf{Reduction over IMI} & \textbf{Reduction over} \texttt{AdpatNN} \\
  \cline{1-7}
  \multirow{6}{*}{Deep1B} & 0.95  & 39.84 ms  & 33.69 ms  & 16.18 ms  & 59\% & 52\% \\ \cline{2-7}
  & 0.96  & 51.21 ms  & 36.65 ms  & 17.77 ms  & 65\% & 52\% \\ \cline{2-7}
  & 0.97  & 69.22 ms  & 44.85 ms  & 21.08 ms  & 70\% & 53\% \\ \cline{2-7}
  & 0.98  & 95.90 ms  & 69.09 ms  & 26.68 ms  & 72\% & 61\% \\ \cline{2-7}
  & 0.99  & 195.45 ms  & 117.94 ms  & 39.61 ms  & 80\% & 66\% \\ \cline{2-7}
  & 0.995 & 275.20 ms & 127.40 ms  & 47.39 ms  & 83\% & 63\% \\ \hline
  \multicolumn{1}{c}{} & \multicolumn{1}{c}{}  & \multicolumn{1}{c}{}  & \multicolumn{1}{c}{}  & \multicolumn{1}{c}{}  & \multicolumn{1}{c}{} & \multicolumn{1}{c}{} \\ \cline{1-7}

  \multirow{6}{*}{Sift1B}
  & 0.95  & 38.69 ms  & 35.53 ms  & 30.93 ms  & 20\% & 13\% \\ \cline{2-7}
  & 0.96  & 46.28 ms     & 39.98 ms  & 32.27 ms  & 30\% & 19\% \\ \cline{2-7}
  & 0.97  & 57.59 ms  & 46.63 ms  & 39.13 ms  & 32\% & 16\% \\ \cline{2-7}
  & 0.98  & 70.78 ms  & 62.79 ms  & 50.71 ms  & 28\% & 19\% \\ \cline{2-7}
  & 0.99  & 117.68 ms  & 98.30 ms  & 69.49 ms  & 41\% & 29\% \\ \cline{2-7}
  & 0.995 & 181.21 ms & 100.73 ms  & 86.71 ms  & 52\% & 14\% \\ \hline
  \end{tabular}
  \label{tab:imi}
  \end{center}
\end{table*}

\subsection{Empirical evaluation with IMI}

Figure~\ref{fig:latency_imi} compares the average end-to-end latency for plain IMI, \texttt{AdaptNN} and \texttt{Tao}. We choose IMI index with OPQ
compression as the baseline, which is one of the state-of-the-art approaches for billion-scale ANN search. We build IMI index with $(2^{14})^2$ = 268,435,456 clusters.


All three methods achieve the highest accuracy, i.e., 0.995 for both Deep1B and Sift1B. We stop at these recall targets because it takes too long to reach 1.0 recall for billion-scale database. Overall, our approach provides up to 1.16x and 2.69x speedup on Sift1B and Deep1B, respectively.

To see the big picture, Table~\ref{tab:imi} lists the latencies for three methods at recall$@$100 targets between 0.95 and 0.995. As the target accuracy increases, the gain of \texttt{Tao} over \texttt{AdaptNN} becomes more and more significant and seems to saturate at the highest recall. The reasons may be two-folds: 1) LID is more informative than features collected by \texttt{AdaptNN}; 2) \texttt{Tao} use only the training data obtained with 1.0 recall whereas \texttt{AdaptNN} is trained at a number of accuracy targets ranging from 0.95 from 0.995. It is worth noting that \texttt{Tao} does not take advantage of any runtime feature as \texttt{AdaptNN} does, and still outperforms its counterpart with the optimal configuration.

\begin{figure}
  \centering
  \includegraphics[scale=0.23]{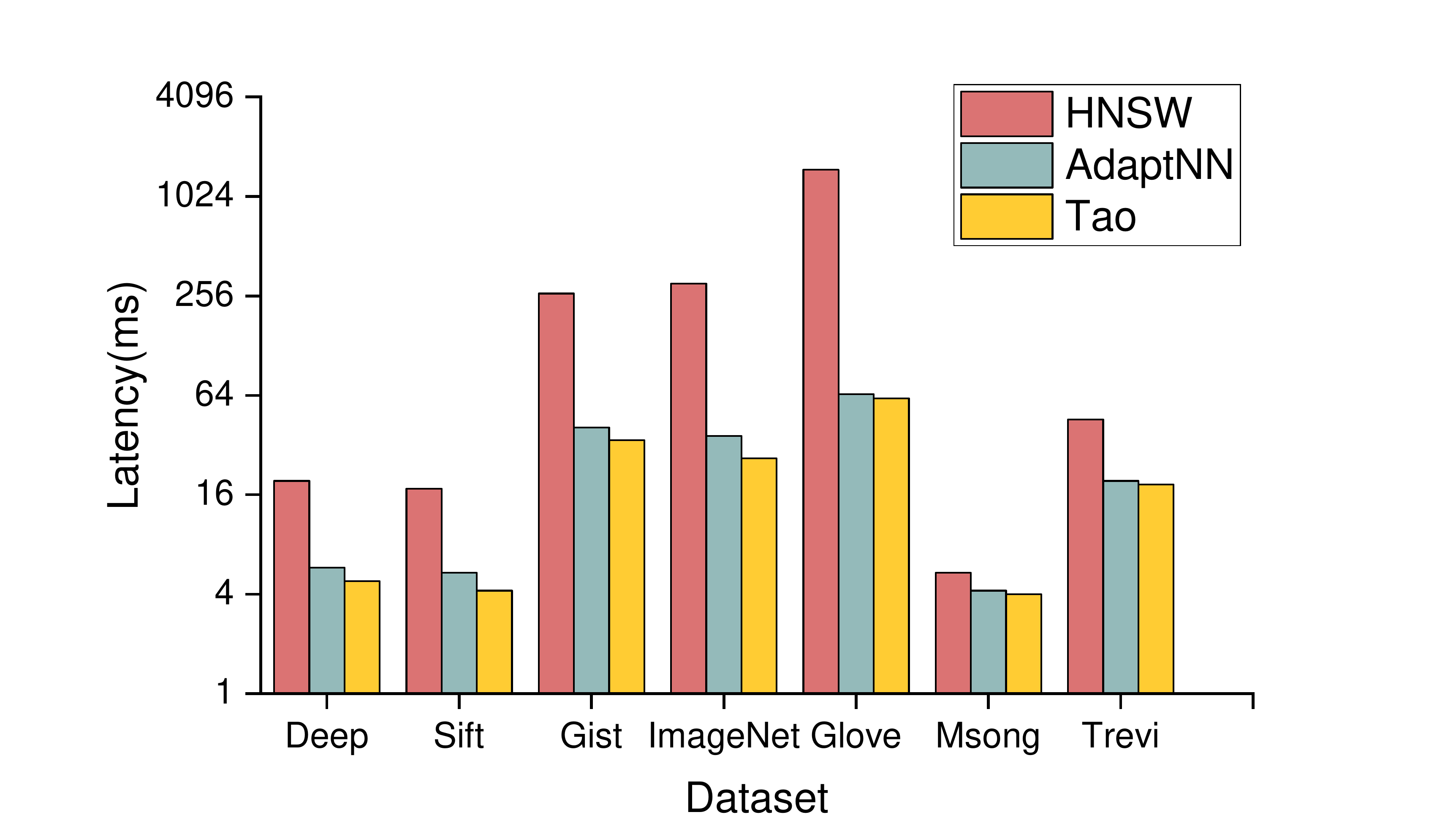}
  \caption{Latency for plain HNSW, \texttt{AdaptNN} and \texttt{Tao}.}
  \label{fig:latency_hnsw}
\end{figure}

\begin{figure}
  \centering
  \includegraphics[scale=0.22]{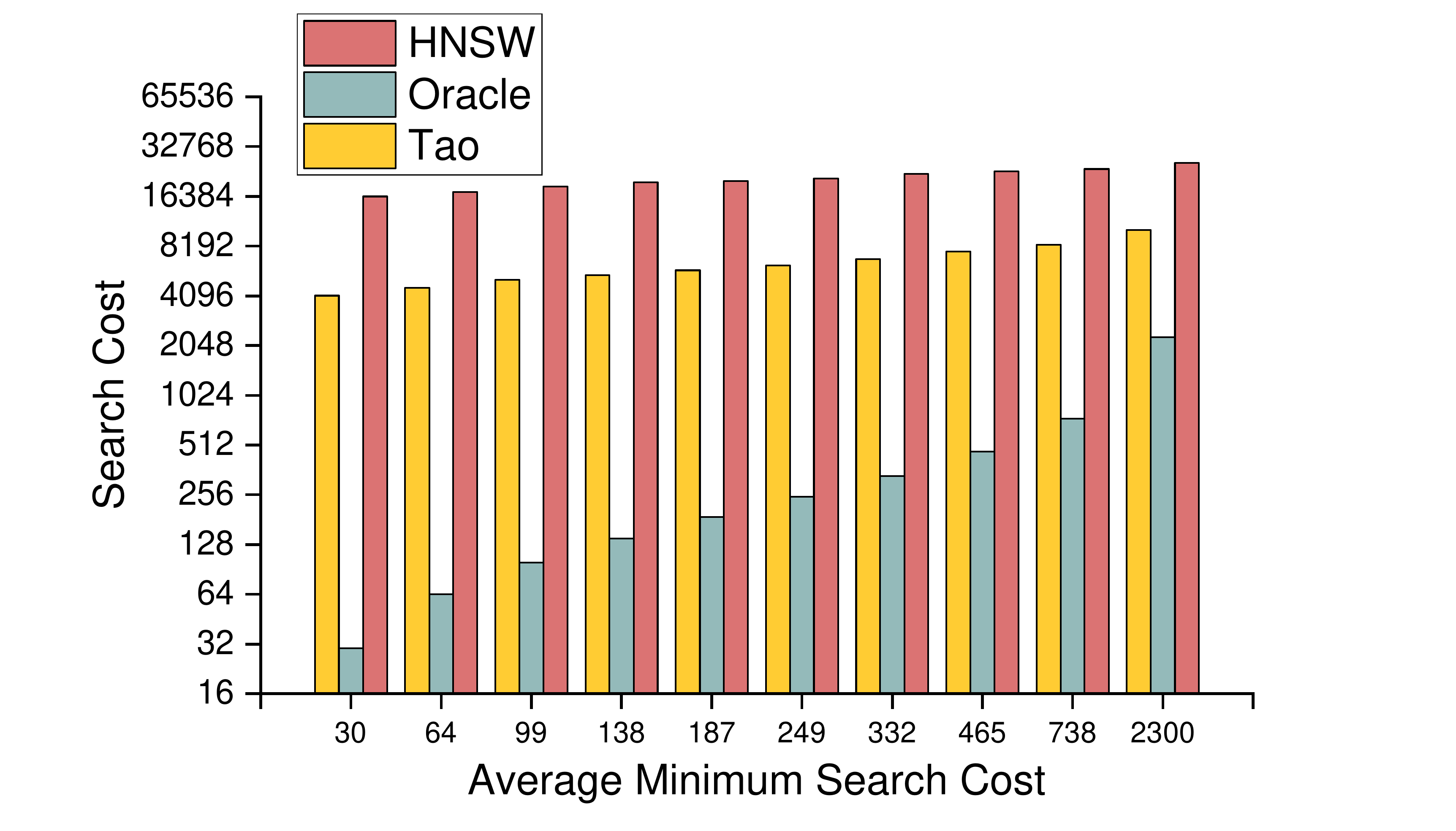}
  \caption{Comparison of search cost for Deep10M.}
  \label{fig:AverageMinumumSearchCost}
\end{figure}

\begin{table}[htbp]
  \caption{Ablation study over Sift10M.}
  \begin{center}
  \begin{tabular}{c|c|c|c}
  \hline
  \textbf{Recall} & \textbf{HNSW} & \texttt{Tao} & \textbf{Real LID}  \\
  \hline \hline
  0.95  & 0.82 ms  & 0.79 ms  & 0.69 ms   \\ \hline
  0.96  & 1.01 ms     & 0.95 ms  & 0.75 ms   \\ \hline
  0.97  & 1.08 ms  & 1.02 ms  & 0.83 ms      \\ \hline
  0.98  & 1.12 ms  & 1.05 ms  & 0.95 ms   \\ \hline
  0.99  & 1.36 ms  & 1.28 ms  & 1.24 ms   \\ \hline
  1.0 & 17.48 ms & 4.28 ms  & 3.81 ms    \\ \hline
  \end{tabular}
  \label{tab:sift_real_lid}
  \end{center}
\end{table}

\subsection{Empirical evaluation with HNSW}

Figure~\ref{fig:latency_hnsw} plots the average end-to-end latencies by comparing plain HNSW, \texttt{AdaptNN} and \texttt{Tao} for highest recalls we have reached.  Because of the graph connectivity issue, it is unable to find the nearest neighbor for a few queries in the reasonable time budget. Hence, for different datasets we stop at different recall targets. Overall, \texttt{Tao} provides consistent speedup over \texttt{AdaptNN} from 1.05x to 1.2x for recall$@$1 measure.


Table~\ref{tab:deep} presents the detailed numbers for different accuracy targets.
As one can see, the performance gaps among three methods are less significant for a relatively low recall, say 0.9 for Glove and 0.95 for the other datasets. With the increase of target recall, both \texttt{AdaptNN} and \texttt{Tao} perform better than the original algorithms thanks to their ability to distinguish `easy' queries from `hard' ones.

Table~\ref{tab:sift_real_lid} lists the results of ablation study over Sift10M -- the performance of \texttt{Tao} after removing the first regression model. We manually calculate LIDs using Equation~\eqref{equ:lid-MLE-estimation} for all queries, and pass them to the second neural network. As one can see, the results are only slightly inferior to those using the predicted LID, which demonstrates the efficacy of the vectors-to-LID mapping.

Figure~\ref{fig:AverageMinumumSearchCost} depicts the average minimum search cost (obtained by the Oracle), the overhead incurred by \texttt{Tao} and fixed configurations in a semi-log plot over Deep10M. By partitioning 10,000 queries, sorted in ascending order of the minimum search cost, into 10 bins evenly, we compute the average of minimum search cost in each bin. As we can see, there exists significant variation in the optimal search cost. As discussed before, HNSW with fixed configuration cannot utilize this fact, thus lead to the largest latency. \texttt{Tao}, instead, is equipped with the power to assign smaller search steps for `easy' queries, by which better performance is obtained. Note that the room for optimization is still giant since what \texttt{Tao} predict is still far away from the Oracle, calling for more efficient features and/or prediction models to shrink this gap.

\subsection{Comparison between \texttt{Tao} and Vector Only Method}

\begin{figure}
  \centering
  \includegraphics[scale=0.23]{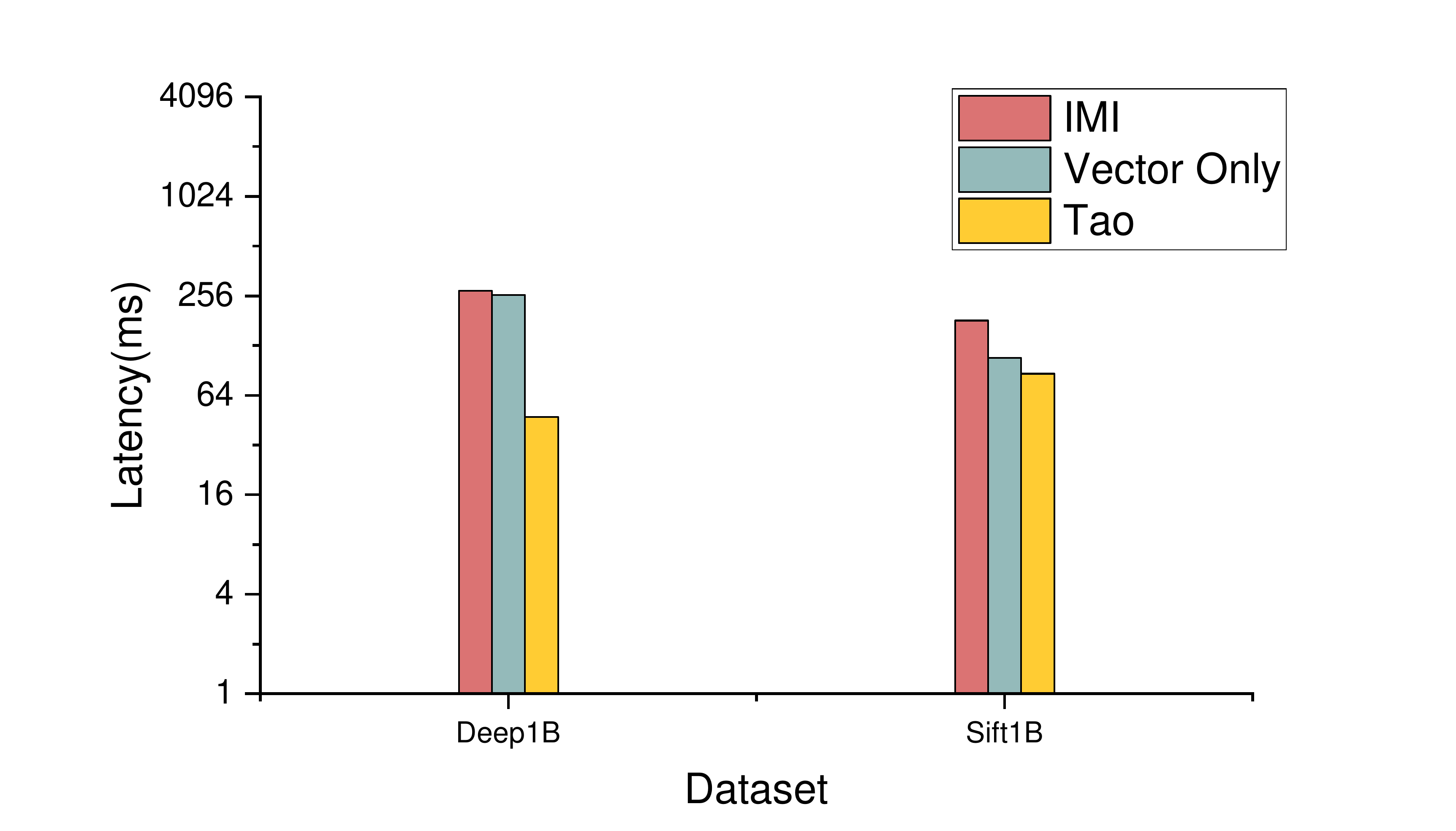}
  \caption{Latency for plain IMI, Vector Only and \texttt{Tao}.}
  \label{fig:vector_only_f_imi}
\end{figure}

\begin{figure}
  \centering
  \includegraphics[scale=0.20]{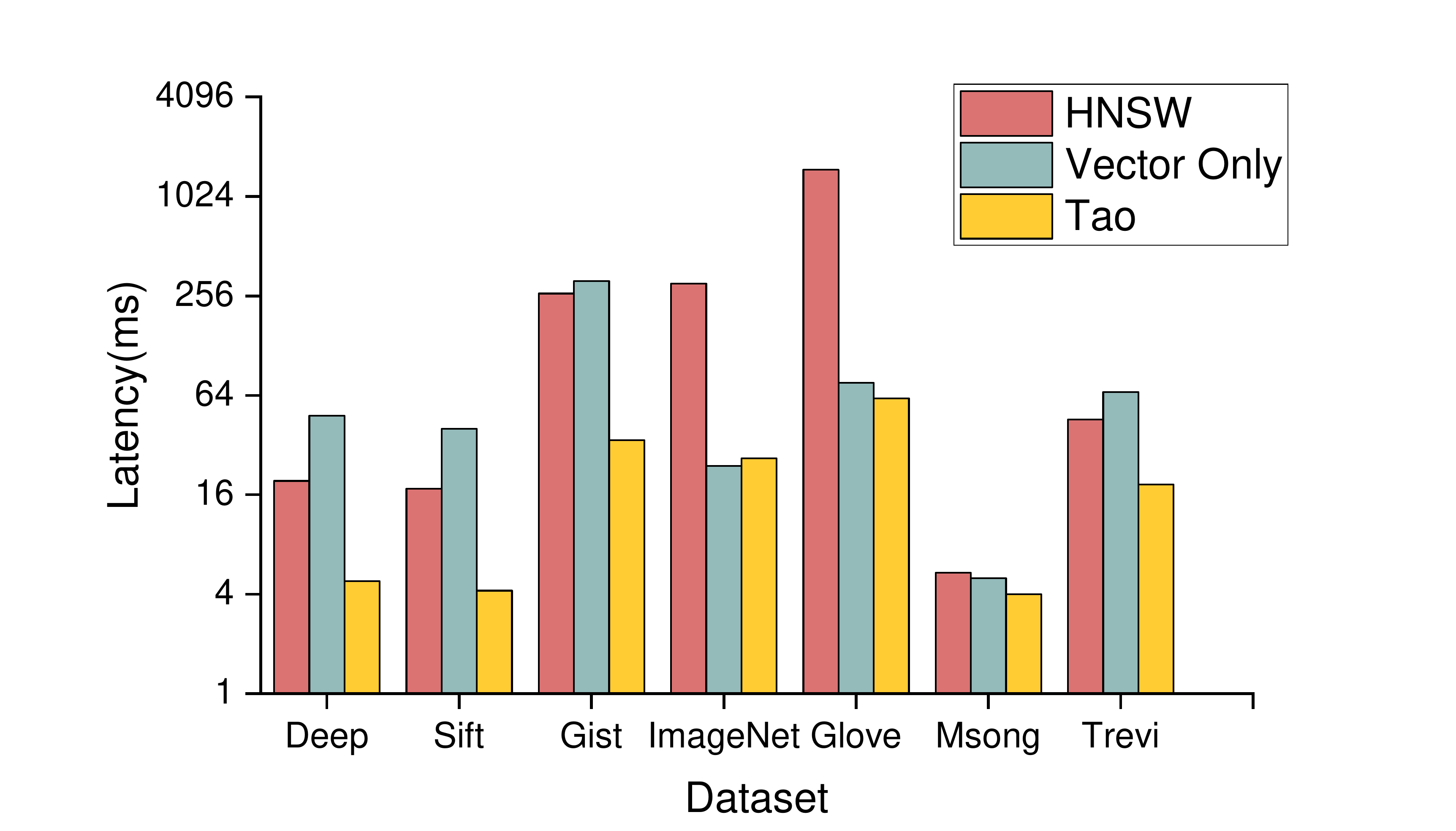}
  \caption{Latency for plain HNSW, Vector only and \texttt{Tao}.}
  \label{fig:vector_only_f_hnsw}
\end{figure}

\begin{table*}[]
  \caption{Latency for plain IMI, Vector Only and \texttt{Tao}.}
  \begin{center}
  \begin{tabular}{c|c|c|c|c|c}
  \hline
  \textbf{Dataset} & \textbf{Recall} & \textbf{IMI} & \texttt{AdaptNN} & \textbf{Vector Only} & \texttt{Tao}     \\ \hline \hline
  Deep1B & 0.995 & 275.20 ms & 127.40 ms & 260.29 ms & \textbf{47.39 ms}  \\ \hline
  Sift1B & 0.995 & 181.21 ms & 100.73 ms & 108.58 ms & \textbf{86.71 ms}  \\ \hline
  \end{tabular}
  \end{center}
  \label{tab:vecotr_only_t_imi}
\end{table*}

\begin{table*}[]
  \caption{Latency for plain HNSW, Vector Only and \texttt{Tao}.}
  \begin{center}
  \begin{tabular}{c|c|c|c|c|c}
  \hline
  \textbf{Dataset} & \textbf{Recall} & \textbf{HNSW} & \texttt{AdaptNN} & \textbf{Vector Only} & \texttt{Tao}     \\ \hline \hline
  Deep10M & 0.999 & 19.47 ms & 5.88 ms & 48.21 ms & \textbf{4.87 ms}  \\ \hline
  Sift10M & 1.0 & 17.48 ms & 5.41 ms & 40.11 ms & \textbf{4.28 ms}  \\ \hline
  Gist   & 0.999  & 264.12 ms & 40.93 ms  & 314.35 ms  & \textbf{34.32 ms} \\ \hline
  ImageNet & 0.998 & 303.34 ms & 36.46 ms & \textbf{24.52} ms  & 26.56 ms \\ \hline
  Glove & 0.9668 & 1486.82 ms & 65.24 ms & 76.07 ms  & \textbf{61.46 ms} \\ \hline
  MSong & 1.0 & 5.48 ms & 4.26 ms  & 4.67 ms & \textbf{4.04 ms}    \\ \hline
  Trevi & 0.997 & 45.85 ms & 19.41 ms & 67.51 ms & \textbf{18.54 ms} \\ \hline
  \end{tabular}
  \end{center}
  \label{tab:vecotr_only_t_hnsw}
\end{table*}

Figure~\ref{fig:vector_only_f_imi} compares the latency of IMI, \texttt{Tao} and VO at the highest recall we achieved, and Table~\ref{tab:vecotr_only_t_imi} lists the detailed numbers including the results of \texttt{AdaptNN}. As we can see: 1) by leveraging the information drawn from query vectors, VO performs better than the plain IMI; 2) without using any runtime features, VO is inferior to \texttt{AdaptNN}, validating the claim made in~\cite{DBLP:conf/sigmod/LiZAH20}; 3) LID does matter and helps to obtain the best latency among all approaches. The experimental results of HNSW are in agreement with IMI as shown in Figure~\ref{fig:vector_only_f_hnsw} and Table~\ref{tab:vecotr_only_t_hnsw}. One interesting observation made is that, in some cases, the latency of VO is even worse than the original HNSW, which is caused by incorrect prediction.

This set of experimental results indicates that 1) the neural network of VO cannot learn the correlation between query vectors and the amount of NN search without the help of LID; 2) the explicit use of LID not only delivers better performance and makes \texttt{Tao} easy to use in practice, it also makes the black box (learning process) more explainable for users compared with \texttt{AdaptNN} and Vector Only method. This advantage is highly desirable nowadays since explainability is key for users to understand conclusions and recommendations of the prediction model.

\section{Related Work}


A large number of ANN search algorithms are available in literature, making this section cannot be exhaustive due to space limitation. The latest benchmarks~\cite{DBLP:journals/tkde/LiZSWLZL20} show that no algorithm dominates the others in all scenarios and each index comes with different tradeoffs in performance, accuracy and space overhead.


%

\subsection{Hashing-based approaches}

For high-dimensional approximate search, the well-known indexing method is locality sensitive hashing (LSH) \cite{DBLP:conf/vldb/GionisIM99}. The main idea is to use a family of locality-sensitive hash functions to hash nearby data points into the same bucket. After the query point goes through the same hash functions, it will get the corresponding bucket number, and only compare the distance between the point in the bucket and the query point. In the end, the $k$ approximate nearest neighbor results that are closest to the query point will be returned. In recent two decades, many LSH-based variants have been proposed, such as E2LSH \cite{DBLP:conf/compgeom/DatarIIM04}, QALSH \cite{DBLP:journals/pvldb/HuangFZFN15}, Multi-Probe LSH \cite{DBLP:conf/vldb/LvJWCL07}, BayesLSH \cite{DBLP:journals/pvldb/SatuluriP12}.


E2LSH, the classical LSH implementations for $\ell_2$ norm, cannot solve $c$-ANN search problem directly. In practice, one has to either assume there exists a ``magical' radius $r$, which can lead to arbitrarily bad outputs, or uses multiple hashing tables tailored for different radii, which may lead to prohibitively large space consumption in indexing. To reduce the storage cost, LSB-Forest~\cite{DBLP:journals/tods/TaoYSK10} and C2LSH~\cite{DBLP:conf/sigmod/GanFFN12} use the so-called virtual rehashing technique, implicitly or explicitly, to avoid building physical hash tables for each search radius.


Based on the idea of query-aware hashing, the two state-of-the-art algorithms QALSH~\cite{DBLP:journals/pvldb/HuangFZFN15} and SRS~\cite{DBLP:journals/pvldb/SunWQZL14} further improve the efficiency over C2LSH by using different index structures and search methods, respectively.
SRS reduces the ANN search problem in a $d$-dimensional space into the range query in an $m$-dimensional projection space (typically $m = 6$), and uses $R$-tree to fulfill this purpose. Recently, Lv et al. proposed VHP, which achieves better efficiency by ingeniously restricting the search space to be the intersection of those of QALSH and SRS. All of the aforementioned LSH algorithms provide probability guarantees on the result quality (recall and/or precision).

Other LSH extensions such as Multi-probe LSH~\cite{DBLP:conf/vldb/LvJWCL07}, SK-LSH~\cite{DBLP:journals/pvldb/LiuCHLS14}, LSH-forest~\cite{DBLP:conf/www/BawaCG05} and Selective hashing~\cite{DBLP:conf/kdd/GaoJOW15} use heuristics to access more plausible buckets or re-organize datasets, and do not ensure any LSH-like theoretical guarantee. A recent paper discussed how to select better hash functions using a deep learning approach~\cite{DBLP:conf/icde/TangWCGCP21}.

\subsection{Quantization-based approaches}
\label{Sec:brief-review-of-quantization-method}
The most influential vector quantization for ANN search is Product Quantization (PQ)~ \cite{jegou2010product}. It seeks to perform a similar dimension reduction to hashing algorithms, but in a way that better retains information about the relative distances between points in the original vector space. Formally, a quantizer is a function $q$ mapping a $d$-dimensional vector $x\in \mathbb{R}^{d}$ to a vector $q(x)\in C = \{c_i; i \in \mathcal{I}\}$, where the index set $\mathcal{I}$ is assumed to be finite: $\mathcal{I}=0 \ldots k-1$. The reproduction values $c_i$ are called centroids. The set $\mathcal{V}_{i}$ of vectors mapped to given index $i$ is referred to as a cell, and defined as

  \begin{displaymath}
    \mathcal{V}_{i} \triangleq\left\{x \in \mathbb{R}^{D}: q(x)=c_{i}\right\}
  \end{displaymath}

The $k$ cells of a quantizer form a partition of $\mathbb{R}^{d}$. So all the vectors lying in the same cell $\mathcal{V}_{i}$ are reconstructed by the same centroid $c_i$. Due to the huge number of samples required and the complexity of learning the quantizer, PQ uses $m$ distinct quantizers to quantize the subvectors separately. An input vector will be divided into $m$ distinct subvectors $u_j$, $1 \leq j \leq m$. The dimension of each subvector is $d^{*} = d/m$. An input vector $x$ is mapped as follows:

\small
\begin{equation*}
    \underbrace{x_{1}, \ldots, x_{d^{*}}}_{u_{1}(x)}, \cdots, \underbrace{x_{d-d^{*}+1}, \ldots, x_{d}}_{u_{m}(x)}
    \rightarrow q_{1}\left(u_{1}(x)\right), \ldots, q_{m}\left(u_{m}(x)\right)
\end{equation*}

\normalsize
where $q_j$ is a low-complexity quantizer associated with the $j^{th}$ subvector. And the codebook is defined as the Cartesian product,

  \begin{displaymath}
    \mathcal{C}=\mathcal{C}_{1} \times \ldots \times \mathcal{C}_{m}
  \end{displaymath}
  and a centroid of this set is the concatenation of centroids of the m subquantizers. All subquantizers have the same finite number $k^{*}$ of reproduction values, the total number of centroids is $k=\left(k^{*}\right)^{m}$.



PQ offers three attractive properties: (1) PQ compresses an input vector into a short code (e.g., 64-bits), which enables it to handle typically one   billion data points in memory; (2) the approximate distance between a raw vector and a compressed PQ code is computed efficiently (the so-called asymmetric distance computation (ADC) and the symmetric distance computation (SDC)), which is a good estimation of the original Euclidean distance; and (3) the data structure and coding algorithm are simple, which allow it to hybridize with other indexing structures.

Original PQ requires examining all vectors during ANN search. To handle billion-scale datasets, advanced indexing structures such as IMI and IVF are developed \cite{babenko2014inverted, jegou2010product}.

\begin{table*}[htbp]
  \caption{Latency for HNSW, \texttt{AdpatNN} and \texttt{Tao}.}
  \begin{center}
  \begin{tabular}{c|c|c|c|c|c|c}
  \hline
  \textbf{Dataset}  &\textbf{Recall} & \textbf{HNSW} & \texttt{AdaptNN} & \texttt{Tao} & \textbf{Reduction over HNSW} & \textbf{Reduction over} \texttt{AdpatNN} \\
  \cline{1-7}
  \multirow{6}{*}{Deep10M} & 0.95  & 0.77 ms  & 0.69 ms  & 0.76 ms  & 1\% & / \\ \cline{2-7}
  & 0.96  & 0.82 ms  & 0.77 ms  & 0.79 ms  & 4\% & / \\ \cline{2-7}
  & 0.97  & 1.01 ms  & 0.91 ms  & 0.95 ms  & 6\% & / \\ \cline{2-7}
  & 0.98  & 1.27 ms  & 1.15 ms  & 1.21 ms   & 6\% & / \\ \cline{2-7}
  & 0.99  & 1.77 ms  & 1.45 ms  & 1.53 ms  & 14\% & / \\ \cline{2-7}
  & 0.999 & 19.47 ms & 5.88 ms  & 4.87 ms  & 75\% & 17\% \\ \hline

  \multicolumn{1}{c}{} & \multicolumn{1}{c}{}  & \multicolumn{1}{c}{}  & \multicolumn{1}{c}{}  & \multicolumn{1}{c}{}  & \multicolumn{1}{c}{} & \multicolumn{1}{c}{} \\ \cline{1-7}

  \multirow{6}{*}{Sift10M}
  & 0.95  & 0.82 ms  & 0.73 ms  & 0.79 ms  & 4\% & / \\ \cline{2-7}
  & 0.96  & 1.01 ms     & 0.89 ms  & 0.95 ms  & 5\% & / \\ \cline{2-7}
  & 0.97  & 1.08 ms  & 0.96 ms  & 1.02 ms  & 6\% & / \\ \cline{2-7}
  & 0.98  & 1.12 ms  & 1.03 ms  & 1.05 ms  & 6\% & / \\ \cline{2-7}
  & 0.99  & 1.36 ms  & 1.24 ms  & 1.28 ms  & 6\% & / \\ \cline{2-7}
  & 1.0 & 17.48 ms & 5.41 ms  & 4.28 ms  & 76\% & 21\% \\ \hline

  \multicolumn{1}{c}{} & \multicolumn{1}{c}{}  & \multicolumn{1}{c}{}  & \multicolumn{1}{c}{}  & \multicolumn{1}{c}{}  & \multicolumn{1}{c}{} & \multicolumn{1}{c}{} \\ \cline{1-7}

  \multirow{6}{*}{Gist}
  & 0.95  & 3.90 ms   & 3.57 ms  & 3.69 ms  & 5\% & / \\ \cline{2-7}
  & 0.96  & 4.86 ms  & 4.23 ms  & 4.56 ms  & 6\% & / \\ \cline{2-7}
  & 0.97  & 6.17 ms  & 5.41 ms  & 5.71 ms  & 7\% & / \\ \cline{2-7}
  & 0.98  & 7.66 ms  & 7.06 ms  & 7.13 ms  & 7\% & / \\ \cline{2-7}
  & 0.99  & 14.40 ms  & 10.92 ms & 10.23 ms & 29\% & 6\% \\ \cline{2-7}
  & 0.999 & 264.12 ms & 40.93 ms & 34.32 ms & 87\% & 16\% \\ \hline

  \multicolumn{1}{c}{} & \multicolumn{1}{c}{}  & \multicolumn{1}{c}{}  & \multicolumn{1}{c}{}  & \multicolumn{1}{c}{}  & \multicolumn{1}{c}{} & \multicolumn{1}{c}{} \\ \cline{1-7}

  \multirow{6}{*}{ImageNet}
  & 0.95  & 1.23 ms  & 0.66 ms  & 0.86 ms  & 30\% & / \\ \cline{2-7}
  & 0.96  & 1.45 ms  & 0.76 ms  & 0.93 ms  & 36\% & / \\ \cline{2-7}
  & 0.97  & 1.79 ms  & 0.91 ms  & 1.14 ms  & 36\% & / \\ \cline{2-7}
  & 0.98  & 2.72 ms  & 1.28 ms  & 1.46 ms  & 46\% & / \\ \cline{2-7}
  & 0.99  & 3.82 ms  & 2.70 ms   & 2.36 ms  & 38\% & 13\% \\ \cline{2-7}
  & 0.998 & 303.34 ms & 36.46 ms & 26.56 ms & 91\% & 27\% \\ \hline

  \multicolumn{1}{c}{} & \multicolumn{1}{c}{}  & \multicolumn{1}{c}{}  & \multicolumn{1}{c}{}  & \multicolumn{1}{c}{}  & \multicolumn{1}{c}{} & \multicolumn{1}{c}{} \\ \cline{1-7}

  \multirow{4}{*}{Glove}
  & 0.9   & 5.08 ms  & 3.79 ms  & 4.49 ms  & 12\% & / \\ \cline{2-7}
  & 0.95  & 94.93 ms & 13.10 ms  & 16.09 ms & 83\% & / \\ \cline{2-7}
  & 0.96  & 335.25 ms & 23.01 ms & 22.12 ms & 93\% & 4\% \\ \cline{2-7}
  & 0.9668 & 1486.82 ms & 65.24 ms & 61.46 ms & 96\% & 6\% \\ \hline

  \multicolumn{1}{c}{} & \multicolumn{1}{c}{}  & \multicolumn{1}{c}{}  & \multicolumn{1}{c}{}  & \multicolumn{1}{c}{}  & \multicolumn{1}{c}{} & \multicolumn{1}{c}{} \\ \cline{1-7}

  \multirow{6}{*}{Msong}
  & 0.95  & 0.60 ms   & 0.53 ms  & 0.58 ms  & 3\% & / \\ \cline{2-7}
  & 0.96  & 0.78 ms  & 0.67 ms  & 0.74 ms  & 5\% & / \\ \cline{2-7}
  & 0.97  & 0.89 ms  & 0.73 ms  & 0.76 ms  & 15\% & / \\ \cline{2-7}
  & 0.98  & 1.01 ms     & 0.65 ms  & 0.83 ms  & 17\% & / \\ \cline{2-7}
  & 0.99  & 1.56 ms  & 0.82 ms  & 1.06 ms  & 32\% & / \\ \cline{2-7}
  & 1.0     & 5.48 ms  & 4.26 ms  & 4.04 ms  & 26\% & 5\% \\ \hline

  \multicolumn{1}{c}{} & \multicolumn{1}{c}{}  & \multicolumn{1}{c}{}  & \multicolumn{1}{c}{}  & \multicolumn{1}{c}{}  & \multicolumn{1}{c}{} & \multicolumn{1}{c}{} \\ \cline{1-7}

  \multirow{6}{*}{Trevi}
  & 0.95  & 2.20 ms   & 2.03 ms  & 1.77 ms  & 20\% & / \\ \cline{2-7}
  & 0.96  & 2.66 ms  & 2.01 ms  & 2.09 ms  & 21\% & / \\ \cline{2-7}
  & 0.97  & 3.81 ms  & 2.64 ms  & 2.88 ms  & 24\% & / \\ \cline{2-7}
  & 0.98  & 6.07 ms  & 3.76 ms  & 4.18 ms  & 31\% & / \\ \cline{2-7}
  & 0.99  & 8.99 ms  & 5.33 ms  & 6.23 ms  & 31\% & / \\ \cline{2-7}
  & 0.997 & 45.85 ms & 19.41 ms & 18.54 ms & 60\% & 4\% \\ \hline

  \end{tabular}
  \label{tab:deep}
  \end{center}
\end{table*}

\subsection{Graph-based approaches}
\label{sec:review-of-graph-based-methods}
The vast majority of graph-based indexing schemes are designed based on three types of theoretical graph models, that is, Delaunay Graphs \cite{DBLP:journals/ijpp/LeeS80}, Relative Neighbor Graphs \cite{DBLP:journals/pr/Toussaint80} and $K$-Nearest Neighbor Graphs \cite{DBLP:conf/wea/ParedesCFN06, DBLP:conf/www/DongCL11}. In practice, however, most search graphs are essentially approximate $K$-nearest neighbor graphs, which mainly distinguish oneself from the other in the edge selection policies.

Inspired by HNSW, several follow-up projects aim to improve the proximity graph-based approach. Douze et al. propose to combine HNSW with quantization \cite{douze2018link}. Navigating Spreading-out Graph (NSG) aims to reduce the graph edge density while keeping the search accuracy \cite{fu2019fast}. SPTAG combines the IVF index and proximity graph for distributed ANN search\cite{ChenW18}. As with HNSW, all of these proximity graph variants employ fixed configurations to perform a fixed amount of graph traversal for all queries. Interested readers are referred to a recent survey for more detailed discussion~\cite{DBLP:journals/corr/abs-2101-12631}, where useful recommendations and guidelines are given for users to choose appropriate graph-based algorithms.

In this subsection, we would like to highlight a few studies focusing on external memory graph-based indices, which may be of great interests to database researchers. This line of research is imperative since the sizes of datasets can easily exceeds capacity of available main memory in a commodity server as more and more vectors are produced with the ubiquity of machine learning techniques.

To the best of our knowledge, Bin Wang et al. first presented ideas to explore slow storage to achieve billion-scale ANNS in a single machine \cite{DBLP:conf/IEEEpact/WangWLSYJV13}. HM-ANN maps the hierrchical design of the graph-based ANNS with memory heterogeneity in HM, thus most accesses will happen in upper layer stored in fast memory to avoid expensive accesses in slow memory without sacrificing accuracy~\cite{DBLP:conf/nips/0015ZL20}. Vamana fully analyzes the construction details of HNSW and NSG, based on which Vamana uses the random neighbor selection strategy of HNSW to increase the flexibility of edge selection. As a result, it is able to construct a high recall SSD index on a single server of 64G \cite{DBLP:conf/nips/SubramanyaDSKK19}. Minjia Zhang et al. presented an ANN solution ZOOM that uses SSD to employ a multi-view approach. Through the efficient routing and optimized distance computation, ZOOM greatly enhance the effectiveness and efficiency while attaining equal or higher accuracy \cite{DBLP:journals/corr/abs-1809-04067}.

\section{Conclusion and Future Work}

In this paper, we propose \texttt{Tao}, a general learning framework for adaptive ANN search in high dimensional spaces. Using LID as an intermediate feature, \texttt{Tao} decouples the prediction phase from query processing, and eliminates the laborious parameter tuning work involved in \texttt{AdaptNN}. Experimental results show that \texttt{Tao} achieves comparable or even better performance compared to \texttt{AdaptNN}. A few interesting problems are worth further investigation. \texttt{Tao} unveil the hidden correlation between query vectors and their LID values, which might deserve further study on its own right. Given a specific ANN search algorithm, how LID affects the difficulty of ANN search in high-dimensional spaces is also an open problem of both theoretical practical importance.

\section*{Acknowledgment}

This work is supported partially by the Fundamental Research Funds for the Central Universities under grant number (No:2232021A-08), NSF of Xinjiang Key Laboratory under grant number (No:2019D04024), Tianjin ``Project + Team" Key Training Project under grant number (No:XC202022). Wei Wang were
supported by ARC DPs 170103710 and 180103411, and D2DCRC DC25002 and DC25003.

\newpage
\small
\bibliographystyle{plain}
\bibliography{egbib}

\end{document}